\documentclass[a4paper, 12pt]{amsart}
\usepackage{amsthm}
\usepackage{amssymb}
\usepackage{amsmath}
\usepackage{amsfonts}
\usepackage[utf8]{inputenc}
\usepackage[dvips,left=1in, right=1in, top=1in, bottom=1in]{geometry}
\usepackage{color}
\usepackage{natbib}
\usepackage{bm}
\usepackage[colorlinks,hyperfootnotes=true]{hyperref} 
\hypersetup{colorlinks=true,raiselinks=true,breaklinks=false,citecolor=blue}

 \newcommand{\domain}{ D}
 
 \newcommand{\brac}[1]{\left ( #1 \right )}
\newcommand{\brak}[1]{\left [ #1 \right ]}
\newcommand{\brat}[1]{\left \{ #1 \right \}}

\newcommand{\reals}{\mathbb{R}}

  \DeclareMathOperator*{\argmax}{arg\,max}

\usepackage{caption,subcaption}
\newtheorem{algo}{Algorithm}
\newtheorem{thm}{Theorem}[section]
\newtheorem{cor}[thm]{Corollary}
\newtheorem{lem}[thm]{Lemma}
\newtheorem{prop}[thm]{Proposition}
\theoremstyle{definition}
\newtheorem{defn}[thm]{Definition}

\newtheorem{ass}[thm]{Assumption}

\theoremstyle{remark}

\newtheorem{exa}[thm]{Example}
\numberwithin{equation}{section}

\newcommand{\PP}{{\mathbb P}}
\newcommand{\QQ}{{\mathbb Q}}

\newcommand{\RR}{{\mathbb R}}
\newcommand{\EE}{{\mathbb E}}

\newcommand{\VV}{\mathbb{V}}

\newcommand{\MM}{{\mathbb M}}

\newcommand{\CC}{{\mathcal C}}
\newcommand{\EN}{{\mathcal E}}
\newcommand{\XX}{{\mathcal X}}

\newcommand{\II}{{\mathcal I}}
\renewcommand{\SS}{\mathbb{S}}
\newcommand{\Cov}{\mathbb{C}\text{ov}}

\newcommand{\tstar}{\theta^*}
\newcommand{\upr}{\underline{\pi}}
\newcommand{\opr}{\overline{\pi}}

\usepackage{graphicx}  

\begin{document}

\title{Optimal Investment and Equilibrium Pricing under Ambiguity}
\author{Michail Anthropelos}
\thanks{Michail Anthropelos would like to thank Scott Robertson, Steven Vanduffel  and Gordan \v{Z}itkovi\'{c}}
\address{Michail Anthropelos, Department of Banking and Financial Management, University of Piraeus.}
\email{anthropel@unipi.gr}
\author{Paul Schneider}
\thanks{Paul Schneider gratefully acknowledges the SNF grant 100018 189086 ``Scenarios'', and expresses his thanks to participants of research seminars at Dublin City University, TU Berlin and Humboldt University Berlin.}
\address{Paul Schneider, USI Lugano and SFI.}
\email{paul.schneider@usi.ch}

\date{\today}

\begin{abstract}
We consider portfolio selection under nonparametric $\alpha$-maxmin ambiguity in the neighbourhood of a reference distribution. We show strict concavity of the portfolio problem under ambiguity aversion. Implied demand functions are nondifferentiable, resemble observed bid-ask spreads, and are consistent with existing parametric limiting participation results under ambiguity. Ambiguity seekers exhibit a discontinuous demand function, implying an empty set of reservation prices. If agents have identical, or sufficiently similar  prior beliefs, the first-best  equilibrium is no trade. Simple conditions yield the existence of a Pareto-efficient second-best equilibrium, implying that heterogeneity in ambiguity preferences is sufficient for mutually beneficial transactions among all else homogeneous traders. These equilibria reconcile many observed phenomena in liquid high-information financial markets, such as liquidity dry-ups, portfolio inertia, and negative risk premia. 
\end{abstract}

\maketitle
\section{Introduction}
Originating from the famous Ellsberg experiment, Knightian uncertainty, or ambiguity,  has become recognized and important factors in decision processes. However, the economic mechanisms to deal with uncertainty, such as worst-case rules, generally do not inherit the simplicity and tractability of the expected utility framework, precluding extensive equilibrium investigations of risk premia on trading strategies, and the separation of premia for uncertainty and risk. In this paper, we extend the literature by developing  equilibrium portfolio trading rules within the $\alpha$-MEU utility\footnote{The $\alpha$-MEU framework, probably first considered by \citet{hurwicz51}, and later formalized by \citet{ghirardatomaccheronimarinacci04} and \citet{frickiijimayaouang20} is built upon a representation of the value attached to an agent's action in terms of worst-case, and best-case expected utility.  We are particularly interested in discontinuities in optimal positions that are hard to investigate through the smooth ambiguity model introduced in \citet{KlibanoffMarinacciMukerji05,KlibanoffMarinacciMukerji09}.} framework formulated in terms of nonparametric discrete distributions in the neighborhood of a nonparametric reference distribution supported on the same state space. Abstracting from distributional model  assumptions allows us to consider economies ranging from binomial, to realistic market scenario modeling. 


Our findings are based on several contributions. Firstly, we find a novel representation of the value function under $\alpha$-maxmin ambiguity as a superposition of expected utility under the prior, or reference, distribution, and the standard deviation of utility under the prior distribution weighted by the degree of ambiguity aversion/preference. This representation is independent of the particular utility function used, and assumptions on initial wealth. 
This representation holds for all values of $\alpha$, and in particular for ambiguity seekers ($\alpha < 1/2$), for which the penalty term becomes positive, meaning that any time an ambiguity seeker trades, the additional term to the expected utility increases.
Our representation thus endogenizes economic \emph{regularization} akin to multiplier utility \citep{hansensargent01,strzalecki11} under ambiguity aversion,   but also  \emph{deregularization} for ambiguity seekers.  Ambiguity preference and allowed distance from the reference distribution  jointly enter the representation as a single parameter $\delta$,  that shows a clear tradeoff between these two concepts. 
At $\alpha=1/2$, the value function collapses into expected utility under the prior distribution, truly justifying the term \emph{ambiguity neutrality} in this case. 

Secondly, we provide a convexity certificate\footnote{We adopt here the convention that concave maximization problems are referred to as convex minimization problems.} to the $\alpha$-maxmin portfolio problem for the case of ambiguity aversion that  allows rapid solutions. It also allows us to sharply quantify trading inertia regions, and monotonicity with respect to $\delta$.  
There is a price region within the support of the asset that is confined by a superposition of the expected payoff and its standard deviation under the reference distribution, that we think of as the bid-ask spread, on which it is optimal not to trade. This result confirms trading inertia statements made previously in \citet{dowwerlang92, caowangzhang05,illedtisch11,easleyohara10} in a static, and in \citet{uppalwang03,maenhout04,trojanivanini04,epsteinschneider08,epsteinschneider10,werner21} in a dynamic setting under multiple prior expected utility. We show that  that random endowments alleviate this staleness.

Ambiguity seekers on the contrary face a highly irregular portfolio optimization problem, the convexity of which depends on the curvature of the utility function, and  the number of states. 
With only two states, two locally convex regions allow nevertheless sufficient tractability, and many of our examples rely on this setting. We provide a fast  algorithm to reach local optima also in settings with more states. 
Generally, ambiguity seekers have no reservation price (their demand function never becomes zero), and are willing both to sell and buy the asset when its price is not too high or too low. Although, a preference for ambiguity  tends to increase the value function with trading, we state simple conditions under which the optimal position of an ambiguity seeker remains finite if the asset price is strictly within the support of the asset payoff. Consistently with the absence of arbitrage, optimal positions become infinite as the asset price approaches the boundaries of the state space. 

Thirdly, based on the first two contributions, we investigate prices, beliefs and ambiguity preference in general equilibrium. We focus on  equilibria through zero net supply market clearing, that in their nontrivial form can only arise from long and short positions.  We can therefore investigate secondary markets, such as stock and options markets, with endogenous trading volume. Most equilibrium models in the ambiguity  literature, such as \citet{caowangzhang05}, are built around fixed positive initial supply and admit only long positions. Among those, the analysis in \citet{beissnerriedel19} operates in complete markets, where prices are given as maximal expectations. \citet{beissnerwerner21} derive equilibrium conditions using quasidifferentiable calculus. Differences of beliefs support first-best equilibria independently of preferences towards ambiguity, in particular also for ambiguity seekers, who typically feature non-convex  preference functionals \citep{rigottishannonstrzalecki08,StrzaleckiWerner11}.  
We prove  existence of a Pareto-optimal second-best market-clearing equilibrium price (the first-best being no trade), that obtains even if investors' beliefs and utility are similar or identical, as long as they are sufficiently diverse with respect to their ambiguity preference.  The second-best equilibrium explains the existence of negative risk premia, and short positions that no ambiguity-averse agent would take in markets where prior distributions can be obtained with high precision.  Surprisingly,  neither the agents'  degree of risk aversion, nor their initial endowment matter for the existence of this equilibrium. To the best of our knowledge, we are the first to provide specific conditions under which ambiguity divergence is a sole source for mutually beneficial trading in a zero-net supply market with homogeneous traders (with respect to their reference prior, utility and initial wealth).

Finally, as an important ingredient to empirical studies, we derive a martingale representation of the equilibrium price that ensures the absence of arbitrage. We provide sharp conditions for non-zero equilibria among ambiguity averse. Ambiguity seekers may be seen as a necessary source for mutually beneficial (second-best) trading. In the case of an economy with homogenous agents, we  derive conditions for equilibria that can be used to reconcile prices with market frictions such as short-selling constraints.


%
%
\bigskip 

\section{Portfolio optimization under $\alpha$-MEU utility}
 
We consider a static one-period world with finite $D$-atomic state space, $|D|\geq 2$, that supports $K$ financial assets with linearly independent payoffs $ S$ that can be associated, for instance, with contingent claims. We also assume the existence of a riskless asset that plays the role of the numeraire. Its payoff is assumed to be equal to zero. Alternatively, the assets are considered discounted, or in forward terms. 
We first start from the position of an investor who considers entering into a trading strategy in the riskless asset and the risky assets with given prices $ \pi$. The investor has initial wealth $w_0$, and a strictly concave utility function $u$ on terminal wealth. Later in our analysis,  prices $ \pi$ will arise endogenously by an equilibrium argument among a finite number of investors.  

The investor has prior subjective probability distribution $\PP_0$ (potentially different from the ones of other investors), assigning strictly positive probability to each atom in $D$. This measure represents her prior beliefs and/or estimations about  future outcomes. Reflecting realistic  market conditions, she is ambiguous about her belief due to estimation error,  or model risk. To deal with this uncertainty, she considers a family of measures in a  neighbourhood of $\PP_0$. Following the literature on ambiguity, the family of such measures is defined as the set of absolute continuous probability measures with respect to $\PP _0$. For utmost tractability, we quantify the neighborhood with quadratic f-divergence, 
\begin{equation}\label{eq:ambiguity_quadratic}
 \XX:=\brat{P:P(s) \geq 0, \,  \forall  s\in \domain,\sum _{s\in \domain}P( s)=1 ,\sum _{ s\in \domain}\frac{P^2( s)}{\PP_0( s)}  \leq c},
\end{equation} 
where $c\geq 1$. Other divergences could be used to measure the distance, such as entropy, that we do not pursue here for simplicity. The parameter $c$ itself could be seen as a measure of uncertainty, in the sense that higher $c$ implies a larger class of potential probability measures, while the case $c=1$ implies no ambiguity ($\XX$ is a singleton). We refer to $\XX$ as the \emph{ambiguity prior}. Note that by design $\XX$ is a convex, bounded and closed,  and hence compact,  set of probability measures on $D$.  In economic terms,  if $\PP _0$ were a martingale distribution, the restrictions on the considered probability measures could be linked to the elimination of  good deals, in the sense that the Sharpe ratio under a measure $P\in\XX$ will not be far away from the Sharpe ratio under measure $\PP_0$ (see also \cite{cerny03}). From a decision-theoretic viewpoint, it is akin to multiplier utility attributed by \citet{epsteinschneider08} to \citet{andersonhansensargent03}. Furthermore, it is a standard form of regularization used in machine learning and data science.  

Investor's ambiguity concerns are translated to the choice of which measure within the set $\XX$ she chooses to assess her decision problem. For a given portfolio choice, there are optimistic and pessimistic probability measures, for instance in the sense that they imply higher or lower expected payoffs. For given random wealth $W$ at the decision horizon, the optimistic measure is determined by the problem
\begin{equation}\label{eq:maximization}
\overline{P}:=\underset{P\in\XX}{\arg\max}\EE_{P}[u(W)],
\end{equation}
while the pessimistic measure solves the problem 
\begin{equation}\label{eq:minimization}
\underline{P}:=\underset{P\in\XX}{\arg\min}\EE_{P}[u(W)].
\end{equation}
Both, the optimistic and the pessimistic problems, are linear in the measure $P\in \XX$, and with the convex and closed prior $\XX$, they constitute well-defined convex optimization problems. In economic terms, it is important to keep in mind  that these worst-case and best-case measures depend on the act of the investor, that together with the payoffs $ S$ determines $W$, and thus  create a feedback loop between action and beliefs.

Under ambiguity concerns and using the parametrization of the optimistic and pessimistic scenarios, the investor's objectives are given by the $\alpha$-MEU problem
\begin{equation}\label{eq:V_a_gen}
V_\alpha (W)=\alpha \min _{P\in \XX}\EE_P[u\left(W\right)]+(1-\alpha)\max _{P\in\XX}\EE_P[u\left(W\right)],
\end{equation}
for some $\alpha\in[0,1]$. 

In particular, when $W$ is created by an investment in $K$ (linearly independent) tradeable risky assets and the single riskless asset, we have $W=W^{ \theta}:=w_0+ \theta\cdot(S- \pi)$, where $\theta$ belongs to an admissible set of investment strategies $\Theta\subseteq\mathbb{R}^K$, made explicit later,
and the objective function becomes   
\begin{equation}\label{eq:V_a}
V_\alpha ( \theta):=\alpha \min _{P\in \XX}\EE_P[u\left(W^{ \theta}\right)]+(1-\alpha)\max _{P\in\XX}\EE_P[u\left(W^{ \theta}\right)].\footnote{Note that we do not assume that $ S$ has non-negative payoffs, which means in particular that the tradeable assets may be derivatives, such as swaps, or combined derivative products, such as bull or bear spreads, and that prices are not necessarily positive. Indeed, our analysis is geared towards derivatives markets where instruments are in zero-net supply, differently from most studies that assume a fixed positive supply of the asset.}
\end{equation}

Preference specification \eqref{eq:V_a_gen} (or \eqref{eq:V_a}) is termed $\alpha$-MEU utility in the literature. It first appeared in \citet{hurwicz51}, and  subsequent instances \citep[for instance][and many others]{epsteinschneider08,frickiijimayaouang20,klibanoff21}. As a special case, \citet{gilboaschmeidler89} multiple prior expected utility emerges with $\alpha=1$. As mentioned above, every act $ \theta$ implies  unique worst-case and best-case beliefs from the convexity of the optimization problems. When evaluating different acts, investors consider at the same time the beliefs that come with them. 

While it is not straightforward ex-ante to imagine ambiguity seeking individuals ($\alpha < 1/2$), the feedback loop between beliefs and actions provides justification. For instance, an agent may have a large  initial endowment in the assets and be afraid of the price impact she may cause when hedging in equilibrium. A preference for uncertainty may be necessary to reconcile her own beliefs with the  optimal position even if her aversion to risk encoded by the felicity function $u$ was high. Indeed, a preference for uncertainty may be necessary to participate in the market at all. Importantly, numerous empirical studies establish the presence of ambiguity seekers and ambiguity averse empirically in experiments \citep{trautmanndekuilen18,kostopoulosmeyeruhr21}.

We use the same letter $s$ both for an atom of  the state space $\domain$, and  also its index by, according to some (partial) ordering. Writing $p_s=P(s)$ and $p_{0s}=\PP _0( s)$ as short-hand, the minimization problem \eqref{eq:minimization} becomes
\begin{equation}\label{eq:min_simple} 
\begin{split}
 \text{minimize }& \sum _{i=s}^{n}u(W_s) p_s \\
 \text{subject to }&p_s\geq 0,\, \sum _{s=1}^{n}p_s=1, \, \sum _{s=1}^{n}\frac{p_s^{2}}{p_{0s}}\leq c.
 \end{split}
\end{equation} 

Optimization problem \eqref{eq:min_simple} is comprised of a convex (linear) objective functional and an intersection of convex cones and subspaces, and can therefore be approached through standard finite-dimensional techniques. Furthermore, for both the maximization \eqref{eq:maximization} and the minimization \eqref{eq:minimization}, a simple condition on the initial probabilities and the divergence parameter $c$ is sufficient to yield internal and economically appealing solutions.

Our first assertion, which we will maintain  for the rest of the paper, greatly improves tractability of the problems to follow. 
\begin{ass}\label{A:condition for c}[Reference prior and minimal reference probability]
\[c<\min_{s\in D}\left\{\frac{1}{1-p_{0s}}\right\}.\]
\end{ass}
Assumption \ref{A:condition for c} is not unreasonable. For instance, a uniform prior for which the minimum in Assumption \ref{A:condition for c} also agrees with its maximum, would require $c<n/(n-1)$. The assumption encodes that probabilities considered in forming beliefs can not become too small and in particular, it is equivalent to having $1-1/c<p_{0s}$ for each $s\in D$. Since  $c$ is close to one for practical purposes, this requirement is not strict.  

Below, we use $\EE_0$, $\SS_0$, $\VV_0$ and $\Cov_0$ to denote  expectation,  standard deviation, variance and covariance under the measure $\PP_0$. Without loss of generality, we may impose that $\underline{p}_s=\overline{p}_s=p_{0s}$ when $W$ is constant (i.e.~$ \theta=0$). Resolving the standard finite-dimensional optimization problem \eqref{eq:min_simple} yields simple and economically  appealing solutions. We emphasize that the following result holds for arbitrary random wealth $W$ (not necessarily  from investments on $ S$), and any strictly concave utility function. 
\begin{prop}\label{P:optimization}[$\alpha$-MEU probability measures]
Let $\PP_0$ and parameter $c\geq 1$ satisfy Assumption \ref{A:condition for c}. Then, $\forall s\in D$, 
\begin{align}\label{eq:p_min}
\underline{p}_s &=p_{0s}\left(1+\frac{\EE_0[u\left(W\right)]-u\left(W_{s}\right)}{\SS_0[u\left(W\right)]}\sqrt{c-1}\right),\\ 
\label{eq:p_max} \overline{p}_s &=p_{0s}\left(1-\frac{\EE_0[u\left(W\right)]-u\left(W_{s}\right)}{\SS_0[u\left(W\right)]}\sqrt{c-1}\right).
\end{align}
In particular, 
\begin{equation}\label{eq:V_a nice}
 V_\alpha (W)= \EE_0[u\left(W\right)]-\sqrt{c-1}(2\alpha-1)\SS_0[u\left(W\right)]. 
\end{equation}
\end{prop}
Hereafter we will use the following notation extensively,
\begin{equation}\label{eq:delta}
\delta:=\sqrt{c-1}(2\alpha-1).
\end{equation}
Note that $\delta$ is positive (negative) for the  ambiguity averse (seeker) and equal to zero for the ambiguity neutral. Also, since $c\leq 2$ from Assumption \ref{A:condition for c}, we readily get that $|\delta|<1$.

There are several noteworthy observations about Proposition \ref{P:optimization} concerning links to the literature and computational features.\footnote{Note that Assumption \ref{A:condition for c} is not necessary for representation \eqref{eq:V_a nice} to hold. If $p_{0s'}$ is lower than $(c-1)/c$, this may lead to the case $\overline{p}_{s'}$ or $\underline{p}_{s'}$ be equal to zero. In this case however, we still have a representation as in Proposition \ref{P:optimization}. In particular, if $p_{0s'}<(c-1)/c$ and $u(W_{s'})=\max_{s\in D}u(W_s)$, then  
\[\overline{p}_{s}=\frac{p_{0s}}{1-p_{0s'}}\left(1-\frac{\tilde{\EE}_0[u\left(W\right)]-u\left(W_{ s}\right)}{\tilde{\SS}_0[u\left(W\right)]}\sqrt{\tilde{d}}\right),\quad \text{for each }s\neq s',\]
where expectation and variance are under measure $\tilde{\PP}(s):=p_{0s}/(1-p_{0s'})$ for each $s\in s'$ and $\tilde{\PP}(s')=0$ and $\tilde{d}:=c(1-p_{0s'})-1$.} 
\begin{itemize}
\item The tilting of the prior distribution into  worst-case and best-case distributions crucially depends only on $\PP_0$-expected utility. Below average ($\PP_0$-expected) utility states  decrease (increase) the worst-case (best-case) distributions. From this structure, objective function \eqref{eq:V_a nice} can be entirely expressed in terms of the prior distribution $\PP_0$, granting an intuitive decomposition into  $\PP_0$-expected utility, and  an endogenous ambiguity preference-weighted (de)regularization  term $\SS_0[u\left(W^{ \theta}\right)]$. It also drastically simplifies any empirical study that aims to investigate the effect of ambiguity aversion to optimal investment choices.  

\item This (de)regularization term's quantitative impact on the objective function is determined from the ambiguity preference. Under ambiguity aversion, expected utility is \emph{regularized}, resembling multiplier utility from \citet{hansensargent01}, while the ambiguity seeker \emph{deregularizes} her objective function. 

\item Let $d:=c-1$. The term $\delta=\sqrt{d}(2\alpha-1)$ in \eqref{eq:delta} on which the (de)regularization rests, emphasizes an indifference between the extent of ambiguity preference/aversion (parametrized through $\alpha$), and the willingness to deviate from the reference distribution $\PP _0$ (parametrized through $d=c-1$). Computationally, this means that the properties of \eqref{eq:V_a nice} depend solely on the sign and magnitude of $\delta$. We exploit this trait in several ways (e.g.~Proposition \ref{P:demand_function} for ambiguity aversion, Proposition \ref{P:seeker} and Algorithm \ref{algo:locally} for ambiguity seekers). In particular, we readily get that the $\alpha$-MEU preference is essentially  a maxmax-preference by decreasing the divergence parameter from $d$ to $\tilde{d}:=d(1-2\alpha)^2$. Note that Assumption \ref{A:condition for c} still holds for $\tilde{d}$.  

\item At $\alpha=1/2$, the preference collapses into $\PP_0$-expected utility and does not depend on the ambiguity prior. It can therefore rightly be associated with ambiguity neutrality. This terminology is not as easily justified in the general case with a convex combination of a best-case, and a worst-case distribution. This result is in line with \citet{beissnerwerner21}, who prove that if the ambiguity prior $\XX$ is symmetric around $h\in \XX$ (meaning that $\XX-h=h-\XX$), then the $\alpha$-MEU preference becomes expected utility under $h$.

\item It follows directly from Jensen's inequality and representation \eqref{eq:V_a nice}, that when prices satisfy $\EE_0[ S]= \pi$, in other words when the prior probability space is such that it prices the assets, the optimal position of an ambiguity averse is zero.  As we will show below, this is not the only price that makes the investor indifferent to trading the asset. On the  contrary, there is an interval of such prices. On the other hand, for ambiguity seekers, there is no reservation price, meaning that for such investors the optimal position is never equal to zero. 

\item Representation \eqref{eq:V_a nice} is precisely of the same functional form 
to define variational preference relations on the space of models as in \citet{lam16} and \citet{oblojwiesel21} with Kullback-Leibler constrained preferences. This similarity is remarkable, since this variational preference relation $\mathbb P \succeq _{KL} \tilde {\mathbb P}$ between two probability measures $\mathbb P$ and $\tilde {\mathbb P}$ is obtained asymptotically, and under \citet{gilboaschmeidler89} multiple prior expected utility ($\alpha=1)$.
\end{itemize}

Representation \eqref{eq:V_a nice}   can also directly link the reduction of the ambiguity concerns to a specific number of moments of the tradeable assets. For instance, when the investor's utility is quadratic, the ambiguity is reduced to the estimation of only two additional moments under $\PP _0$. In particular, consider for simplicity the case of a single asset, and impose time-additive quasi-linear quadratic utility, 
\[u(x,s)=x-\frac{1}{2\gamma}\left(1-\gamma s\right)^2,\qquad\forall x\in\reals\text{ and }s\in D,\] 
where $\gamma>0$ stands for the investor's risk aversion.  Representation \eqref{eq:V_a nice} then reads
\[V_\alpha(\theta)=w_0-\theta \pi-\frac{1}{2\gamma}\EE_0[\left(1-\gamma \theta S\right)^2]-\delta\frac{1}{2\gamma}\SS_0\left[\left(1-\gamma \theta S\right)^2\right],\]
which is equivalent to 
\[\hat{V}_\alpha(\theta)=\theta(\EE_0[S]-\pi)-\frac{\gamma}{2}\EE _0[(\theta S)^2]-\delta\frac{1}{2\gamma}\SS_0\left[\left(1-\gamma \theta  S\right)^2\right].\]
If an investor is ambiguity-neutral $(\alpha=1/2)$, her preferences are quadratic, and only two moments under the initial probability measure $\PP _0$ have to be considered. On the other hand, if $\alpha\neq 1/2$, the extra term that reflects the ambiguity concerns includes the third and the fourth central moment of the asset's payoff. For this, we calculate that
\begin{align*}
\hat{V}_\alpha(\theta)&=\theta(\EE_0[S]-\pi)-\frac{\gamma\theta}{2}\EE_0[S^2]\\
&-\delta\theta\SS_0[S]\sqrt{1+\frac{\gamma^2\theta^2}{4}\frac{\EE_0[S^4]-\EE_0^2[S^2]}{\VV_0[S]}-\gamma\theta\frac{\EE_0[S^3]-\EE_0[S^2]\EE_0[S]}{\VV_0[S]}}.
\end{align*} 
It emerges from the above formula, that although ambiguity refers to the probability of any state, investors with quadratic preferences need to include only two additional moments (i.e., skewness and kurtosis) in order to cope with   uncertainty. In fact, ambiguity aversion is translated into aversion to negative skewness and   kurtosis in this case. We also emphasize this direct representation of ambiguity through the additional moments holds, even if the number of possible states are arbitrarily large. This feature provides a link to aversion to higher-order risk  in decision theory \citep{ebert13} without ambiguity, 
and distributionally robust optimization  with moment ambiguity sets \citep{nie2021distributionally,hansenetal21}.

Importantly, the notion of certainty equivalent alters under ambiguity too. Within the present setup, the (buyer) certainty equivalent  is the compensation that makes the investor indifferent between buying $ \theta$ units of the asset or the current status quo. More precisely (and with a slight abuse of notation), $\nu( \theta)$ is the certainty equivalent for $ \theta$ units of $ S$ if $V_\alpha(w_0+ \theta \cdot S-\nu( \theta))=u(w_0)$. For example, under exponential utility we readily get that the buyer's certainty equivalent is
\begin{equation}\label{eq:indifferenceexponentialutil}
 \nu( \theta)=-\frac{1}{\gamma}\log\left(\EE_0[e^{-\gamma \theta\cdot  S}]+\delta\SS_0[e^{-\gamma \theta \cdot  S}]\right),\qquad \theta\in\reals^K.
\end{equation}

This means that ambiguity aversion decreases the buyer's certainty equivalent (the investor bids lower prices for risky assets), while an ambiguity seeker is willing to bid higher prices for any number of units $ \theta$. Similarly, we can show that the seller's certainty equivalent increases (decreases) with ambiguity aversion (preference).

\smallskip

Assumption \ref{A:condition for c} proves valuable also for the next results, including bounds on worst-case and best-case utility realizations, stochastic dominance, and strict concavity of the objective functions \eqref{eq:V_a_gen} and \eqref{eq:V_a}. For these results, we recall that $d=c-1\geq 0$.
\begin{lem}[Bounds on the state space]\label{L:condition}
Let $d>0$. Assumption \ref{A:condition for c} implies that 
\begin{align*}
\max_{s\in D}u(W_{s}) &<\EE_0[u(W)]+\frac{\SS_0[u(W)]}{\sqrt{d}},\text{ and}\\
\min_{s\in D}u(W_{s}) &>\EE_0[u(W)]-\frac{\SS_0[u(W)]}{\sqrt{d}}.
\end{align*}
\end{lem}
Another immediate consequence of Assumption \ref{A:condition for c} is an important ingredient for portfolio optimization.   
\begin{prop}[Stochastic dominance]\label{P:increasing}
Under Assumption \ref{A:condition for c}, objective $V_\alpha$ is strictly increasing, in the sense that if random wealth $W_1$ stochastically dominates random $W_2$ under probability measure $\PP_0$, then $V_\alpha(W_1)>V_\alpha(W_2)$.
\end{prop}

We now return to the wealth generated by the investment on the risky and riskless assets, that is $W=w_0+ \theta\cdot( S- \pi)$, for $\theta\in\Theta$, where $\Theta\subseteq\mathbb{R}^K$ is the convex set of admissible strategies. We will consider utility functions with domain equal to entire real line, and also utility functions that are finite only for non-negative wealth. Since we do not impose any additional constraints on the investment choices (short selling, or borrowing),  the set of admissible strategies is $\Theta=\mathbb{R}^K$ in the case of a real-valued domain. On the other hand, when utility admits only positive wealth, the admissible strategies for an investor with initial (positive) wealth $w_0$ is the easily implemented box constraint
\begin{equation}\label{eq:Theta_w0}
\Theta_{w_0}(\pi ):=\{ \theta\in\mathbb{R}^K\,:\, w_0+ \theta\cdot(s- \pi)>0,\,\forall   s\in D\},
\end{equation}
where we recall that $ \pi$ stands for the price vector of risky assets. This constraint is important to rule out pathologies that can arise in asset pricing through negative wealth.  For instance,  quadratic utility assigns the same felicity to positive or negative wealth,  leading to paradoxa such as negative pricing kernels \citep{cerny03}.

For the rest of this section, we study the demand function problem, or in other words the optimal investment strategy of an investor with ambiguity concerns, starting with the case of  ambiguity aversion. 

\subsection{Demand function under ambiguity aversion} 

The following proposition is at the heart of our study, as it ensures the existence of a unique solution, if $\alpha\geq 1/2$, that can be obtained through convex optimization techniques independently of the number $K$ of assets.

\begin{prop}[Strict concavity]\label{P:concavity}
Let $u$ be a strictly concave function and suppose that Assumption \ref{A:condition for c} holds. Then, if $\alpha\geq 1/2$, the objective function $V_{\alpha}:\Theta\mapsto \mathbb{R}$ is strictly concave.
\end{prop}

The above Proposition \ref{P:concavity} promises great tractability for portfolio problems involving ambiguity seekers. 
Further properties of $V_\alpha$ are to be explored, such as its behavior at zero, since the price at which the optimal position is zero yields the reservation price (for prices higher than this, the ambiguity averse investor sells and for lower prices she buys). However, representation \eqref{eq:V_a nice} implies that $V_\alpha$ is \textit{not} differentiable at zero, as we substantiate in Appendix \ref{sec:diffatzero}, and therefore does not admit the standard way of calculating reservation prices at $ \theta =  0$. Below we nevertheless derive  demand functions,  that can nevertheless be developed.\footnote{In a dynamic context, \citet{marimonwerner21} provide remedies for dealing with non-differentiable value functions.  As we are particularly interested in non-zero portfolio weights, our analysis can be performed without further methodological contributions.}

Having proved the concavity of $V_\alpha$ for $\alpha\geq 1/2$, we now examine the optimal position in the tradeable asset. In other words, we are interested to see the first-order condition (FOC) that gives the investor's demand function at each level of price $ \pi$. From the above analysis,  the demand function ${\theta}^*$ solves
\begin{equation}\label{eq:FOC}
\EE_{0}[u'(W^*)( S- \pi)]=\delta \frac{\EE_0\left[(u(W^*)-\EE_0[u(W^*)])u'(W^*)( S- \pi)\right]}{\SS_0[u(W^*)]},
\end{equation}
where $W^*:=w_0+{\theta}^*\cdot( S-\pi)$. Note that \eqref{eq:FOC} can be expressed as 
\begin{equation}\label{eq:FOC_measure}
\frac{\EE_{\PP^*}[u'(W^*) S]}{\EE_{\PP^*}[u'(W^*)]}= \pi,\quad\text{ where}\quad p_s^*=p_{0s}\left(1-\delta\frac{u(W^*_s)-\EE_0[u(W^*)]}{\SS_0[u(W^*)]}\right).
\end{equation}
The above relations show that the investor solving her investment problem implies the existence of a martingale probability measure $\QQ \sim \PP ^*$ with 
\begin{equation}\label{eq:martingalemeasure}
q _s=p _s^*\brac{\frac{u'(W^*)}{\EE_{\PP^*}[u'(W^*)]}},
\end{equation}
that arises endogenously from her optimal trading strategy. From the fundamental theorem of asset pricing, the presence of a martingale measure also implies the absence of arbitrage opportunities. This martingale  representation arises analogously to the one in  expected utility, and also under ambiguity \citet[][Proposition 6]{beissnerwerner21}.  
Before summarizing the main properties of the demand function $\tstar( \pi )$ for a single asset,  we introduce some additional notation (that refers to all $\alpha \in[0,1]$),
\begin{align}
\opr&=\EE_0[S]+\vert\delta\vert\SS_0[S], \label{eq:opr}\\
\upr&=\EE_0[S]-\vert\delta\vert\SS_0[S]. \label{eq:upr}
\end{align}

With $\underline{S}:=\min_{s\in D}$ and $\overline{S}:=\max_{s\in D}$, it follows from Assumption \ref{A:condition for c} that 
\[\underline{S}<\upr<\opr<\overline{S}.\]
This is because $\upr$ can be written as $\EE_{\underline{P}}[S]$, where $\underline{p}_s:=p_{0s}\left(1+\vert\delta\vert(\EE_0[S]-S_s)/\SS_0[S]\right)$.  Note that Lemma \ref{L:condition} implies that $\underline{P}$ is a well-defined probability measure, equivalent to $\PP$. Similarly, we may write $\opr$ as $\EE_{\overline{P}}[S]$, where $\overline{p}_s:=p_{0s}\left(1-\vert\delta\vert(\EE_0[S]-S_s)/\SS_0[S]\right)$, and again by Lemma \ref{L:condition} we get that $\overline{P}$ is a well-defined probability measure, equivalent to $\PP_0$. 
In fact, we show below that $[\upr,\opr]$ is the reservation interval of an ambiguity-averse investor. 

With this notation, we now consider the case of a single asset and $\Theta=\mathbb{R}$ or $\Theta=\Theta_{w_0}(\pi )$. The upper and lower bounds of the admissible set are denoted by $\underline{\Theta}$ and $\overline{\Theta}$, respectively, and in particular, we allow for two different types of portfolio restrictions,  
\begin{equation}\label{eq:bounds of Theta} 
\begin{split}
&\text{For }\Theta=\mathbb{R},\ \underline{\Theta}:=-\infty \text{ and }\overline{\Theta}:=+\infty.\\
&\text{For }\Theta=\Theta_{w_0}(\pi ),\ \underline{\Theta}:=-\frac{w_0}{\overline{S}-\pi} \text{ and }\overline{\Theta}:=\frac{w_0}{\pi-\underline{S}}.
\end{split}
\end{equation}

We first state  that Assumption \ref{A:condition for c} implies that the optimal demands are in the interior of the investor's admissible set. For this, we need to impose the so-called Inada conditions for the utility functions.
\begin{ass}\label{A:Inada}[Inada conditions]
For the utility functions $u$ defined on the whole real line we assume $\underset{x\downarrow-\infty}{\lim}u'(x)=+\infty$ and for those defined only for positive wealth we assume $\underset{x\downarrow 0}{\lim}u'(x)=+\infty$. Also for both cases, $\underset{x\uparrow+\infty}{\lim}u'(x)=0$.
\end{ass}

Note the following results hold for all $\alpha\in[0,1]$, but are constrained to the one-asset case for simplicity. We also use the following simplified notation 
\[V'_\alpha(\overline{\Theta}):=\underset{\theta\uparrow\overline{\Theta}}{\lim}V'_\alpha(\theta)\qquad\text{ and }\qquad V'_\alpha(\underline{\Theta}):=\underset{\theta\downarrow\underline{\Theta}}{\lim}V'_\alpha(\theta).\]
\begin{lem}\label{lem:large theta}
Let Assumption \ref{A:condition for c} hold, assume that $K=1$, and consider a strictly increasing and strictly concave utility function $u$ that satisfies Assumption \ref{A:Inada}. Then, for all $\alpha\in[0,1]$ and for all prices $\pi\in(\underline{S},\overline{S})$ it holds that 
\begin{equation}\label{eq:large theta}
V'_\alpha(\overline{\Theta})=-\infty\quad\text{ and }\quad V'_\alpha(\underline{\Theta})=+\infty.
\end{equation}
\end{lem}

The  above Lemma \ref{lem:large theta} is relevant for the discussion of  optimal demand under ambiguity. As the validity of the statement is independent of ambiguity preferences, we see that even in the case of an ambiguity seeker, the optimal demand is never unbounded when the price of the risky asset is within the non-arbitrage bounds $(\underline{S},\overline{S})$. Indeed, these limits imply in particular that $V'_{\alpha}$ is decreasing for large $\theta$ and increasing for low $\theta$. Hence, a portfolio allocation  at the boundaries can not be optimal (note that the second limit is similar to the related Inada condition usually, and herein, imposed for utility functions). While this result is expected for an ambiguity-averse agent, who we would intuit to hesitate to take larger positions in the risky asset, it is less obvious for the ambiguity seeker. Indeed, although an ambiguity seeker can increase her utility when she trades (see representation \eqref{eq:V_a nice} for $\alpha<1/2$), the concavity of her utility function together with the standing Assumption \ref{A:condition for c} implies that the optimal demand positions are within the admissible set of investments.

We are now ready to examine the demand function of an ambiguity averse investor. 

\begin{prop}[Demand function and reservation interval]\label{P:demand_function}
Let Assumption \ref{A:condition for c} hold, assume that $\alpha\in[1/2,1]$, $K=1$ and consider a strictly concave utility function $u$ that satisfies Assumption \ref{A:Inada}. Then, for the solution of \eqref{eq:FOC} the following hold:
\begin{itemize}
\item[i.] $\tstar(\pi)$ is unique and finite for all $\pi\in\left(\underline{S},\overline{S}\right)$.
\item[ii.] $\tstar(\pi)$ is continuous on $\left(\underline{S},\overline{S}\right)$, equal to zero on $[\upr,\opr]$ and $\tstar(\pi)>0$ for prices $\pi\in\left(\underline{S},\upr\right)$  and $\tstar(\pi)<0$ on $\left(\opr, \overline{S}\right)$.
\item[iii.] The optimal demand for lower and upper no-arbitrage prices are 
\[\underset{\pi\downarrow\underline{S}}{\lim}\,\tstar(\pi)=\overline{\Theta}\quad\text{ and }\quad\underset{\pi\uparrow\overline{S}}{\lim}\,\tstar(\pi)=\underline{\Theta}.\]
\end{itemize}
\end{prop}

Nonparticipation due to ambiguity is a well-established possibility in the literature. In the context of multiple prior expected utility ($\alpha=1$), \citet{dowwerlang92}, and \citet{epsteinschneider07} with recursive multiple priors in a dynamic learning context, establish the intuitive portfolio inertia result that risk-averse (risk-neutral) investors will engage in trading if and only if risk premia are positive.   With reference-dependent preferences, \citet{guasonimeireles-rodrigues20} show that non-participation can already arise due to loss aversion. 
Proposition \ref{P:demand_function},  ii. confirms this notion also for the $\alpha$-MEU framework, but it also is more explicit about the precise intervals that make trading optimal. Furthermore, it confirms the intuition, that as prices approach the no-arbitrage bounds, investors go ``all in''.

\begin{figure}[ht]
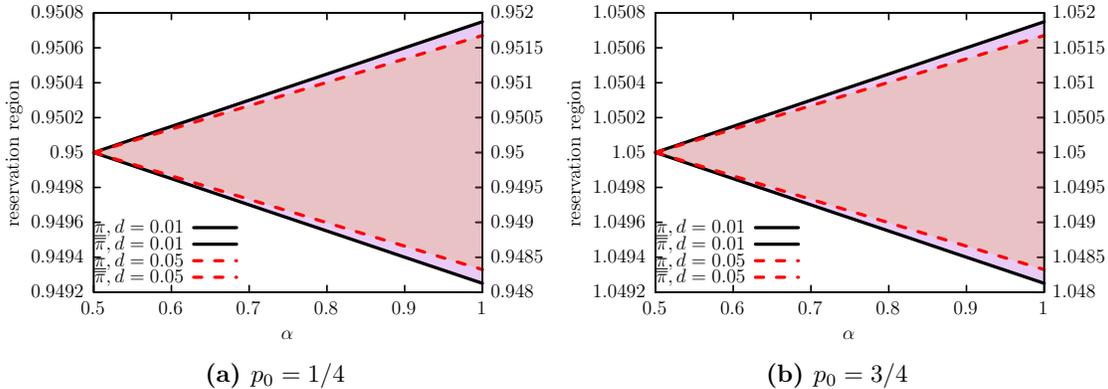

\begin{subfigure}[t]{0.49\textwidth}
 \scalebox{0.65}{\input{reservationlo}}
  \caption{\label{fig:reserlo}{$p_0=1/4$}}  
\end{subfigure}
\begin{subfigure}[t]{0.49\textwidth} 
 \scalebox{0.65}{\input{reservationhi}}
  \caption{\label{fig:reserhi}{$p_0=3/4$}}  
\end{subfigure}
\caption{\label{fig:reser}Reservation region. The plots show the reservation region in a binomial model with state space $S_g=1.1, S_b=0.9$ for a power-utility agent with risk aversion $\gamma=2$ as a function of her ambiguity preference $\alpha$ for an asset with price $p=1$. The reservation region for $d=0.01$ are plotted on the left y axis, for $d=0.05$ on the right y axis. The left (right) panel shows the preference function with  prior distribution $\PP _0(S_g)=1/4$ ($\PP_0(S_b)=3/4$). }  
\end{figure} 

Figure \ref{fig:reser} shows  that the reservation region is increasing in ambiguity aversion, as expected. It is also relatively small quantitatively compared to the price of the asset in the binomial case, well within the magnitude of bid asks spreads prevailing in many financial markets. 

The next result goes in a similar direction, showing that the magnitude of the position taken in the one-asset case decreases with ambiguity aversion. This is a non-trivial and important result, as equally as intuitively one could imagine  situations in which investors hedge against uncertainty, demanding larger positions with increased uncertainty aversion. 


\begin{prop}[Condition for decreasing demand demand w.r.t. ambiguity]\label{P:comparative_statics_delta}
Let Assumption \ref{A:condition for c} hold, assume that $\alpha\in(1/2,1]$, $K=1$ and consider a strictly increasing and strictly concave utility function $u$ that satisfies Assumption \ref{A:Inada}. If the relative risk aversion of $u$ is bounded from below by one, then $\theta^*$ as a function of $\alpha$ is increasing for prices in $(\opr,\bar{S})$ and decreasing in prices $(\underline{S},\upr)$.  
\end{prop}
An interesting aspect of Proposition \ref{P:comparative_statics_delta} above is the importance of the coefficient of relative risk aversion, which is however only sufficient for the statement to hold. With power utility, for instance, log utility represents the minimum degree of risk aversion for the result to hold. This interplay between risk and ambiguity aversion may be an empirically interesting feature. From \eqref{eq:derivative of theta wrt alpha} in the proof for the statement, we see also that the same relation holds for $d$, that determines the size of the ambiguity ball around $\PP_0$. Finally, from the same equation, where we identify  the second derivative of $V_\alpha$ as the numerator, the statement is also true (locally) for an ambiguity seeker at her optimal demand.


\subsection{The effect of  random endowments}
An insightful implication of ambiguity aversion arises from  existing risk exposure (random endowments). The portfolio inertia results from above rely  on the absence of such random endowments. There are many instances that can be modeled with random endowments, however, such as   an initial position in the tradeable asset (a position to be hedged), or some unhedgeable existing background risk. Nevertheless, the analysis above remains almost unchanged. 

Representation \eqref{eq:V_a nice} does not depend on the form of the terminal wealth $W$, while the concavity of $V_\alpha$ (for $\alpha\geq 1/2$) still holds, with the only difference being that $W^{ \theta}=w_0+\theta(S-\pi)+\EN$, for some random variable $\EN$. The most important difference occurs with  portfolio inertia, that vanishes when $\EN$ and $S$ are linearly independent. In fact, it is straightforward to read from \eqref{eq:derivative} that $V_\alpha$ is everywhere differentiable,  implying that under linearly independent random endowment, the reservation interval becomes a singleton. This is intuitive, since an unhedgeable existing risk keeps the investor away from certainty, which is the very mechanism that creates  inertia under ambiguity aversion. In particular, in the multi-asset case, multivariate  demand functions reduce to demand functions in the single asset case with random endowments. In the two-asset case, for instance,   the inertia interval becomes an inertia ``cross'', rather than an inertia sphere. 

The situation is slightly different when $\EN=\theta_1 S$ for some $\theta_1\in\RR$. In this case, the reservation interval is a singleton, but $V_\alpha$ is not differentiable at $-\theta_1$ (the proof is similar to one of the second item of Proposition \ref{P:demand_function}). In fact, 
\[
\underset{\theta\uparrow -\theta_1}{\lim}V'_{\alpha}(\theta)=u'(w_0)\left(\opr-\pi\right)\quad\text{and}\quad \underset{\theta\downarrow -\theta_1}{\lim}V'_{\alpha}(\theta)=u'(w_0)\left(\upr-\pi\right),
\]
which implies that for prices within the interval $(\upr=\EE_0[S]-\vert\delta\vert\SS_0[S],\opr=\EE_0[S]+\vert\delta\vert\SS_0[S])$, the optimal demand is constant and equal to $-\theta_1$. In other words, when the investor's endowment is hedgeable, the need of an 
ambiguity averse investor to hedge (i.e., to return to certainty) is more pressing, which makes her willing to pay more for the position, than in the case of no ambiguity.

\subsection{Demand function for ambiguity seekers}\label{sec:ambiguity seeking}
For ambiguity-seeking investors,   quantitative and qualitative properties of optimal demand change drastically compared to the case of ambiguity aversion. This section focuses on this case and works out these differences. We pursue this question, since empirical evidence amply points towards the presence of ambiguity seekers in the market place \citep{trautmanndekuilen18,kostopoulosmeyeruhr21}.

Recall that Proposition \ref{P:optimization}, and in particular representation \eqref{eq:V_a nice}, hold even when $\alpha<1/2$. From this property, we can deduce that the  investor's objective function increases not only when  expected utility (under $\PP _0$) increases, but also when utility's standard deviation increases. This implies that an ambiguity-seeker investor has an additional motive to trade more, all else equal.

As  representation \eqref{eq:V_a nice} consolidates the parameters of ambiguity aversion and divergence into one single parameter $\delta=\sqrt{d}(2\alpha-1)$,  the optimization problem for ambiguity seekers can be written as a max-max problem, meaning that $\alpha$ is set to be zero, and $d$ is set to $\delta ^2$. While Proposition \ref{P:concavity} guarantees that function $V_\alpha$ is strictly concave when $\alpha\geq 1/2$, this is not always true if $\alpha< 1/2$, however. In particular, we readily get from Lemma \ref{lem:concavity} that  $\MM(X)=\EE_0[X]-\delta\SS_0[S]$ remains monotone, but becomes even (locally) convex when $\delta$ is negative. In  view of the proof of Proposition \ref{P:concavity}, and since $V_\alpha(\mathbb{ \theta})=\MM(u( \theta))$, we get that the objective becomes (locally) concave, if the concavity of the utility function sufficiently compensates for the convexity of $\MM$. More precisely, when $K=1$ (the case of one asset), we calculate that for all $\theta\neq 0$,
\begin{equation}\label{eq:horrorderivative}
\begin{split}
V_\alpha''(\theta)&=\EE_0\left[u''(W^\theta)(S-\pi)^2\right]
-\delta\frac{\Cov_0\left(u(W^\theta),u''(W^\theta)(S-\pi)^2\right)}{\SS_0[u(W^\theta)]} \\
&-\frac{\delta}{\SS_0[u(W^\theta)]} \left(\SS_0^2[u'(W^\theta)(S-\pi)]-\frac{\Cov_0^2\left(u(W^\theta),u'(W^\theta)(S-\pi)\right)}{\SS_0^2[u(W^\theta)]}\right).
\end{split}
\end{equation}
The formula above implies that the (local) concavity of the objective function $V_\alpha$  arises from the  number of states, the values of $S$ at the terminal time, and the particular choice of the utility function. To see this, consider the case of the binomial model, where the number of states is the smallest possible, and we do have local concavity for every choice of strictly (local) concave utility (see Subsection \ref{sub:binomial} below), whereas when utility approaches linearity (the investor becomes risk neutral), $V_\alpha$ tends to become (locally) convex due to the convexity of $\MM$.

\begin{figure}
 \input{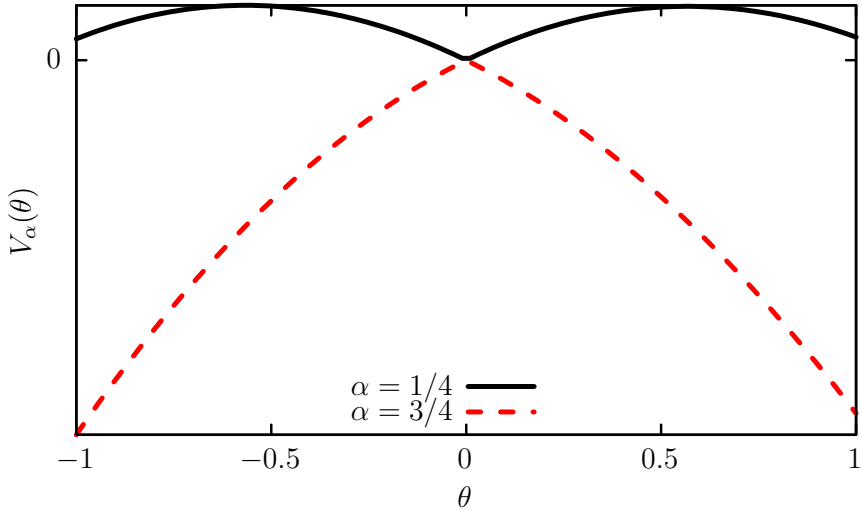} 
 \caption{\label{fig:utilexample}Ambiguity preference and utility function. This figure shows $V_\alpha(\theta)$ for the ambiguity seeker ($\alpha=1/4$) and the ambiguity averse ($\alpha=3/4$) for a log utility specification,  $c=1.01$ and a single asset with two-atomic state space. The prior distribution $\PP _0$ puts the same probability on both states $S_g=1.1$ and $S_b=0.9$.}
\end{figure}

\begin{figure}
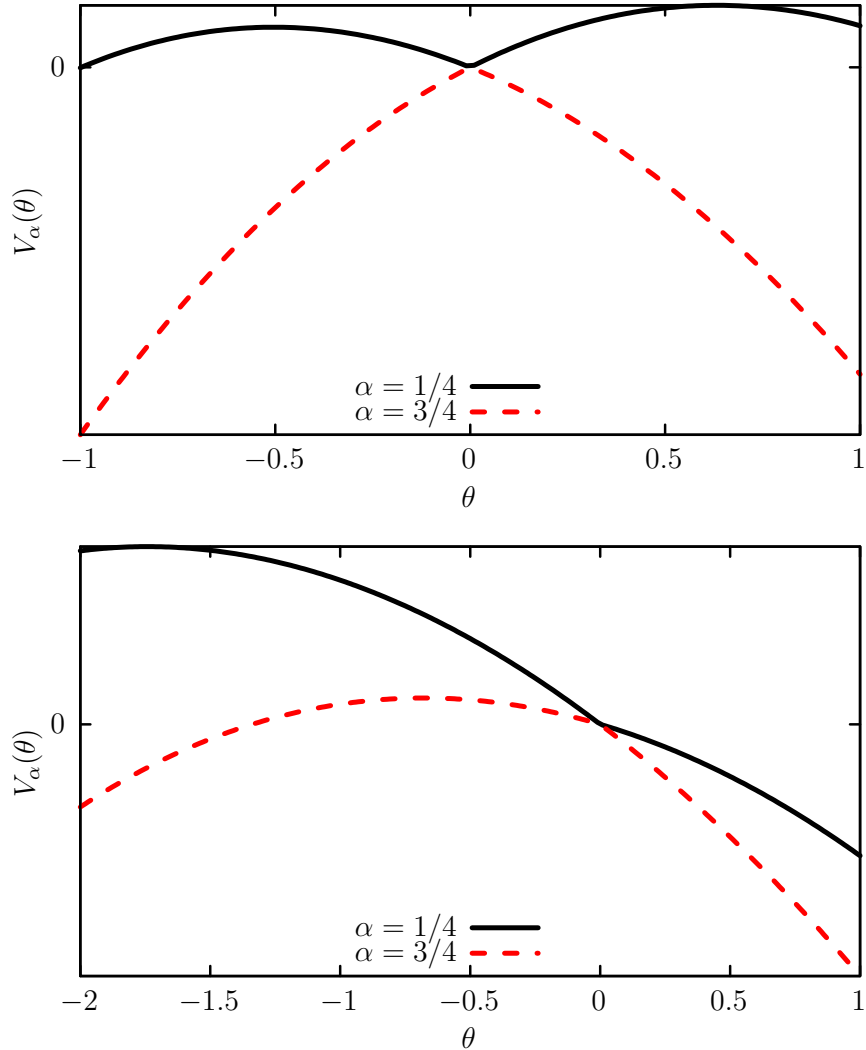

 \input{utilexample_downprice}  
 \input{utilexample_upprice} 
 \caption{\label{fig:utilexample_}Ambiguity preference and utility function. This figure shows $V_\alpha(\theta)$ for the ambiguity seeker ($\alpha=1/4$) and the ambiguity averse ($\alpha=3/4$), for a log utility specification, $c=1.01$  and a single asset with two-atomic state space. The top (bottom) panel shows the preference function with  prior distribution $\PP _0(S_g)=1/4$ ($\PP_0(S_b)=3/4$) with  $S_g=1.1$ and $S_b=0.9$.}
\end{figure}

Figure \ref{fig:utilexample} shows an example of $V_\alpha $ as a function of $\theta$ in a symmetric binomial economy in the single asset case for an ambiguity seeker $(\alpha=1/4)$, and an ambiguity averse $(\alpha=3/4)$ where $\mathbb E _0 \brak{S}=\pi$. The preference of the ambiguity averse peaks at 
$\theta=0$, making it optimal for her not to trade. The preference of the ambiguity seeker features two locally concave regions away from zero that make her indifferent to buying or selling (see Proposition \ref{p:binomial_seeker} in this context). 
For neither ambiguity seekers, nor  ambiguity averse, the  preference function is differentiable at zero. Figure \ref{fig:utilexample_} features a skewed economy, where the price $\pi$ is above (below) its $\PP _0$ expectation. In the top panel, $\EE_0[S]<\pi< \overline{\pi}$, such that the ambiguity seeker has two locally concave regions with their respective local maxima. The bottom panel has $\pi < \underline{\pi}<\EE_0[S]$, such that the ambiguity seeker has but one globally optimal position. 
In both economies, the preference functions of the ambiguity averse (seeker) are concave (locally concave). Both the ambiguity seeker and the averse want to take the same side in trading the asset, long or short. This indicates that  the \citet{dowwerlang92} risk premium result can be extended also to the case of an ambiguity seeker.

Based on Figures \ref{fig:utilexample} and \ref{fig:utilexample_}, we conjecture that $V_\alpha$ is not globally concave (as in the case of ambiguity aversion), but locally concave on $(-\infty,0)$ and on $(0,\infty)$. Two types of strategies exist to cope with this problem. The first is a condition on the degree of risk aversion that renders  \eqref{eq:horrorderivative} concave. The second regularizes the portfolio weight $ \theta$ with some norm, to provide enough convexity for the problem to become well behaved. 

Reading from  representation \eqref{eq:V_a nice}, we can tell whether an ambiguity seeker has a locally optimal positive and/or negative position, even if her objective is not concave for positive and negative positions (as in Figures \ref{fig:utilexample} and \ref{fig:utilexample_}). This is a very interesting problem because, an ambiguity seeker has an extra motive to trade on both sides, as long as the price is not too low or too high. We first verify that for prices lower than $\underline{\pi}$ and higher than $\overline{\pi}$, the ambiguity seeker is willing to trade only on  one side.\footnote{Recall that $\opr$ and $\upr$ are defined in \eqref{eq:opr} and \eqref{eq:upr}, respectively.} We then state the conditions that guarantee the existence of  demand  that is locally optimal  for both, long and short positions,  and discuss its implications afterwards.   

%
%

\begin{prop}\label{P:seeker}
Let Assumption \ref{A:condition for c} hold, assume that $\alpha\in[0,1/2)$, $K=1$ and consider a strictly increasing and strictly concave utility function $u$ that satisfies Assumption \ref{A:Inada}. Then, 
\begin{itemize}
\item[i.] For $\pi\geq\opr$, $\underset{\theta\geq 0}{\argmax}\,V_\alpha(\theta)=0$. 
\item[ii.] For $\pi\leq\upr$, $\underset{\theta\leq 0}{\argmax}\,V_\alpha(\theta)=0$.
\item[iii.] For $\pi\in[\EE_0[S],\opr)$, there exists at least one negative and one positive locally optimal finite demand. 
\item[iv.] For $\pi\in(\upr,\EE_0[S]]$, there exists at least one positive and one negative locally optimal finite demand. 
\end{itemize}  
\end{prop}

From i. and ii. of the above Proposition  \ref{P:seeker} above, if the price of the tradeable asset is too high (resp.~too low), the ambiguity seeker is willing  to sell only (resp.~too buy only). For prices within the interval $(\upr,\opr)$, however,  an ambiguity seeker desires to take both long and short positions, as we see in iii. and iv. above. In other words, her value function has non-zero local maxima on both sides of the trade,  and her demand function is never equal to zero, as illustrated already by Figure  \ref{fig:utilexample_}.

%

Based on these results, we readily can deduce that the demand function is discontinuous at zero, which essentially means that the set of reservation prices for an ambiguity seeker is empty when the risky asset's price lies in $(\opr,\upr)$. We state this insightful result in the following corollary. 

\begin{cor}\label{cor:discontinuity}
Depart from the assumptions made in Proposition \ref{P:seeker}. Then, 
\[\underset{\pi\downarrow\EE_0[S]}{\lim}\theta^*(\pi)\neq\underset{\pi\uparrow\EE_0[S]}{\lim}\theta^*(\pi), \]
where $\theta^*(\pi)$ denotes optimal demands for prices close to $\EE_0[S]$. In addition, when $\pi\in(\opr,\upr)$, there is no locally optimal demand equal to zero.   
\end{cor}

Financial economics is oftentimes concerned with finding motives that induce trading into financial markets, such as hedging, and production-based positions. Corollary \ref{cor:discontinuity} highlights that ambiguity preference alone may be a sufficient mechanism to generate liquidity, as the ambiguity seeker has no reservation price, or interval. 

Furthermore, similarly to the case of an ambiguity averse trader (Proposition \ref{P:comparative_statics_delta}), any local optimal demand of an ambiguity seeker is increasing (in absolute value) as a function of the ambiguity parameter $\alpha$. 

\begin{prop}\label{P:comparative_statics_delta_seeker}
Let Assumption \ref{A:condition for c} hold, assume that $\alpha\in[0,1/2)$, $K=1$ and consider a strictly increasing and strictly concave utility function $u$ that satisfies Assumption \ref{A:Inada}. If the relative risk aversion of $u$ is bounded from below by one, then any local optimal demand $\theta^*$ as a function of $\alpha$ is increasing for prices in $(\opr,\bar{S})$ and decreasing in prices $(\underline{S},\upr)$.  
\end{prop}

\subsection{An efficient algorithm for ambiguity seekers' optimal demand}
Having worked out the existence of local optima in the one-asset case in  Proposition \ref{P:seeker}, we develop in this section a numerical algorithm that yields the locally optimal portfolio weights (for positive or negative positions). The algorithm is designed for pure ambiguity seekers $\alpha = 0$, but can be applied without loss of generality also for any $\alpha \in (0,1/2)$, as 
Representation \eqref{eq:V_a nice} allows us to rewrite the 
$\alpha$-MEU preferences, by  changing the divergence parameter from $d$ to $\tilde{d}:=d(1-2\alpha)^2$ (recall that Assumption \ref{A:condition for c} holds for $\tilde{d}$ too). Hence, computationally, any ambiguity seeker with $\alpha \in [0,1/2)$ is equivalent  to a pure ambiguity seeker (a max-max optimizer) with decreased divergence ball. 

She obtains her portfolio position as
\begin{equation*}
\theta ^*=\argmax _{\theta\in\Theta}\max _{P\in\XX}\EE_P[u\left(W^{\theta}\right)].
\end{equation*}
As we have seen in the previous sections, this problem is not jointly concave in $\theta$ and $P$. However, the presence of the maximization operators,  and in connection with Proposition \ref{P:seeker}, suggests an algorithm to obtain the locally optimal demand as follows. 
\begin{algo}[Locally optimal portfolios, $\alpha<1/2$]\label{algo:locally}
For fixed $\varepsilon >0$, set $\tilde{d}:=d(1-2\alpha)^2$
 \begin{enumerate}
  \item Fix $\theta _0$ and set $i=0$
  \item Set $P_i=\argmax _{P\in \mathcal X}\EE_P[u\left(W^{ \theta _i}\right)]$
  \item Set $\theta _{i+1}=\argmax _{\theta}\EE_{P_i}[u\left(W^{\theta }\right)]$
  \item If $| \theta _{i+1}- \theta _{i}|\leq \varepsilon$ terminate. Otherwise increment $i$ by one and return to (2).
 \end{enumerate} 
\end{algo}
Note that the optimization problems in (2) and (3) are well-behaved, since the solution to (2) is given by \eqref{eq:p_max}, and (3) is the standard expected utility problem with a unique solution,  provided $u$ is strictly concave. Since the sequence of maximizations ensures that the value function is non-decreasing, Algorithm \ref{algo:locally} is guaranteed to converge to a local maximum, as Lemma \ref{lem:large theta} guarantees that the optimal portfolio weight is bounded. 

Importantly, Algorithm \ref{algo:locally} is applicable to both high-dimensional state spaces, as well as multi-asset settings. In practice, one can start the algorithm in the one-asset case from $\theta _0<0$, and in parallel from $\theta _0>0$, and compare the utility drawn from the locally optimal solutions from both instances. In the multi-asset case, this imposes limits on the number of assets that can be processed,   as all combinations of negative/positive portfolio starting weights would need to be considered. 
 
\begin{figure}
 \input{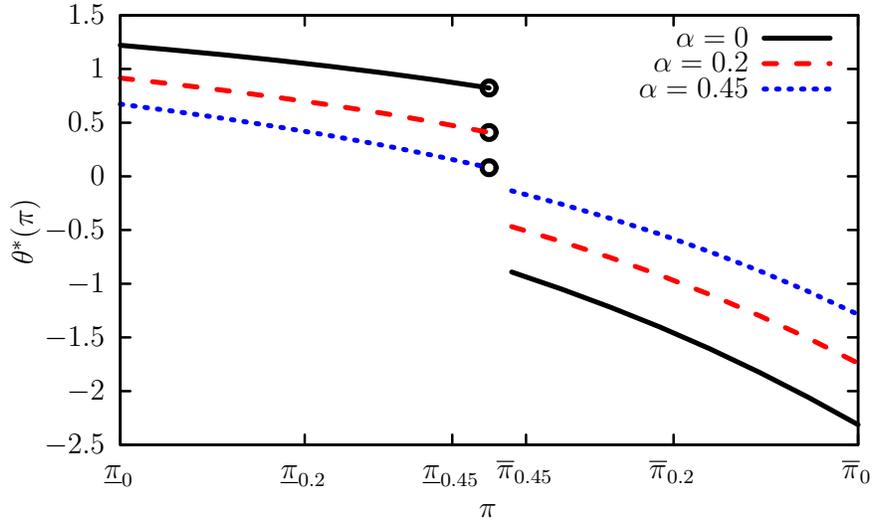}
  \caption{\label{fig:threestate}Ambiguity seeker's demand function in trinomial model. The figure shows the ambiguity seeker's optimal demand for different degrees of ambiguity preference computed from Algorithm \ref{algo:locally}. The parameters generating the picture are $d=0.01$, with the asset taking values $0.5,1,$ and $1.1 $ with probabilities $0.05,0.7$, and $0.25$, respectively, such that $\EE_0[S]=1$. The black dot indicates the position taken at $\EE_0[S]=1$.}
\end{figure}             

Figure \ref{fig:threestate} shows the optimal demand computed from Algorithm \ref{algo:locally} with a three-dimensional state space. As expected the magnitude of the positions desired increases in $\alpha$, with the side of the trade taken changing discontinuously at a price level that yields zero risk premium. Equally as intuitively, the position for $\alpha=0.45$ starts to resemble the optimal demand of an expected utility maximizer.

\subsection{Binomial Example}\label{sub:binomial} 

\begin{figure}[ht]
 \input{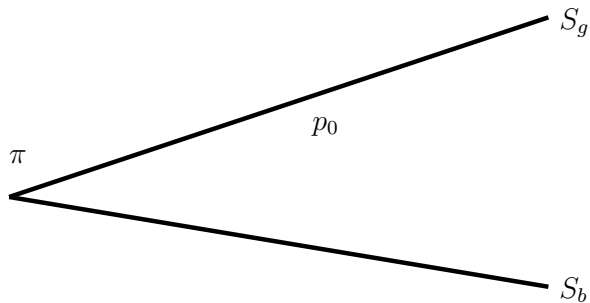}  
 \caption{\label{fig:binomial}Binomial model. The figure shows the binomial model with  state space $S_g>S_b$ and given prior probability $p_0$, as well as price $\pi$.}
\end{figure}

Although the binomial case may appear simplistic, it still contains all the important aspects of the model, especially for  $\alpha < 1/2$. In particular, as we shown below, in a binomial economy the value function $V_\alpha$ is locally concave for positive and for negative positions on the risky asset, respectively,  as long as the utility function is strictly concave. The below Figure \ref{fig:binomial} shows two possible payoff states, $S_g$ and $S_b$ and  prior probability $p_0$ on state $S_g$.

From $|D|=2$, and  maintaining Assumption \ref{A:condition for c}, we readily calculate that,
\begin{equation}\label{eq:V_bin}
V_\alpha(W)=p_0u(W_g)+(1-p_0)u(W_b)-\sqrt{d}(2\alpha -1)\sqrt{p_0(1-p_0)}|u(W_g)-u(W_b)|,
\end{equation}
where $p_0$ is the probability of risky asset's increase and $W_g$ and $W_b$ is the value of investor's wealth at the good,  and the bad  event, respectively. In particular, when $W=w_0+\theta(S-\pi)$, we have that $W_g=w_0+\theta(S_g-\pi)$. Note that $V_\alpha$ is not differentiable at zero due to the absolute value in \eqref{eq:V_bin} above. Furthermore,  the objective function differs from positive to negative $\theta's$. Rewriting \eqref{eq:V_bin} in these terms,
\begin{equation}\label{eq:V_bin_new}
\begin{split}
V_\alpha(\theta)&=V^+_\alpha(\theta):=\EE_+[u(W^{ \theta})],\text{ for }\theta\geq 0\text{, and } \\ V_\alpha(\theta)&=V^-_\alpha(\theta):=\EE_{-}[u(W^{ \theta})],\text{ for }\theta< 0,
\end{split}
\end{equation}
where $\EE_+$ and $\EE_-$ denote expectations under probability measures
\begin{align}
 P_+&=(p_0-\sqrt{d}(2\alpha -1)\sqrt{p_0(1-p_0)},1-p_0+\sqrt{d}(2\alpha -1)\sqrt{p_0(1-p_0)}),\text{ and } \label{eq:pplus}\\
 P_-&=(p_0+\sqrt{d}(2\alpha -1)\sqrt{p_0(1-p_0)},1-p_0-\sqrt{d}(2\alpha -1)\sqrt{p_0(1-p_0)}),\label{eq:pminus}
\end{align}
respectively.\footnote{The tractability of the binomial case arises due to the expected utility representation \eqref{eq:pplus} and \eqref{eq:pminus}.  
 \citet{beissnerwerner21} show that any binomial model under $\alpha$-MEU has this property. In fact, in our model, the optimism and pessimism probability measures depend only on the sign of the investment position on the risky asset, and are independent of its size.} 
Also, for prices $\pi \in(S_b,S_g)$, the FOC is written as 
\begin{align*}
\frac{u'(W_g)}{u'(W_b)}&=\frac{(\pi-S_b)(1-p_0+\delta\sqrt{p_0(1-p_0)})}{(S_g-\pi)(p_0-\delta\sqrt{p_0(1-p_0)})},\quad\text{ for }\theta>0, \text{ and }\\
\frac{u'(W_g)}{u'(W_b)}&=\frac{(\pi-S_b)(1-p_0-\delta\sqrt{p_0(1-p_0)})}{(S_g-\pi)(p_0+\delta\sqrt{p_0(1-p_0)})},\quad\text{ for }\theta<0.
\end{align*}
Furthermore, in addition to Proposition \ref{P:concavity} that deals with $\alpha\geq 1/2$, in the binomial case we also have that $V_\alpha$ is strictly concave on $(0,\infty)$ and on $(-\infty,0)$, even when $\alpha\in[0,1/2)$. In particular, for $\theta>0$ 
\begin{equation*}
 \begin{split}
V''_\alpha(\theta)&=u''(W_g)(S_g-\pi)^2(p_0+|\delta|\sqrt{p_0(1-p_0)}) \\
&+u''(W_b)(S_b-\pi)^2(1-p_0-|\delta|\sqrt{p_0(1-p_0)})> 0
\end{split}
\end{equation*}
and for $\theta<0$,
\begin{equation*}
 \begin{split}
V''_\alpha(\theta)&=u''(W_g)(S_g-\pi)^2(p_0-|\delta|\sqrt{p_0(1-p_0)})\\
&+u''(W_b)(S_b-\pi)^2(1-p_0+|\delta|\sqrt{p_0(1-p_0)})> 0,
\end{split}
\end{equation*}
since Assumption \ref{A:condition for c} guarantees that $p_0\pm|\delta|\sqrt{p_0(1-p_0)},(1-p_0)\pm|\delta|\sqrt{p_0(1-p_0)}>0$. 

\begin{prop}\label{p:binomial_seeker}
Let $|D|=2$, $K=1$ and $\alpha<1/2$. The following holds for any strictly concave utility function $u$.
\begin{itemize}
\item[i.] $\underset{\theta\geq 0}{\argmax}V_\alpha(\theta)$ is unique, finite,  and strictly positive if and only if $\underline{S}<\pi<\opr$, 
\item[ii.] $\underset{\theta\leq 0}{\argmax}V_\alpha(\theta)$ is unique,  finite, and strictly negative if and only if $\upr<\pi<\overline{S}$, 
\end{itemize}
where $\opr$ and $\upr$ are defined in \eqref{eq:opr} and \eqref{eq:upr}, respectively. 
\end{prop}

The above proposition illustrates that the notion of reservation price does not exist for an ambiguity seeker. In other words, when $\alpha<1/2$, the demand function never becomes zero \emph{as it is never optimal to stay at certainty}. More precisely, for any price $\pi$ within the interval $(\upr,\opr)$, the optimal demand may be positive or negative. Indeed, \citet{dowwerlang92} show that $\tstar$ is positive  for $p>\EE_0[S]$, and negative for $p<\EE_0[S]$ under non-additive expected utility and/or multiple prior expected utility. Furthermore, as we show below, the demand function is discontinuous at $\EE_0[S]$.


\begin{figure}[ht]
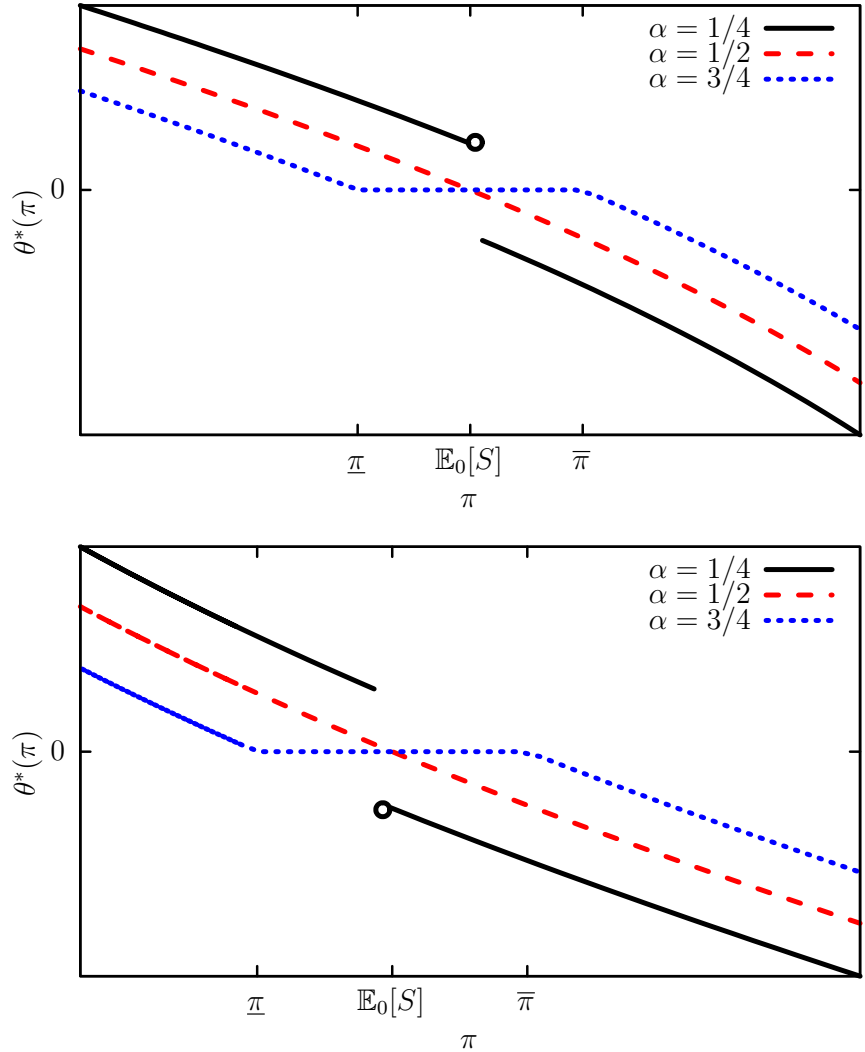
 
 \input{demandhip0}       
 \input{demandlop0}       
 \caption{\label{fig:demandfunctions}Ambiguity preference and utility function. This figure shows the optimal demand $\theta ^*(p)$ for the ambiguity seeker ($\alpha=1/4$), the ambiguity neutral ($\alpha=1/2$), and the ambiguity averse ($\alpha=3/4$) for a power utility specification ($\gamma = 2$) and $c=1.01$. The top panel shows the optimal demand for $p_0=3/4$, while the bottom panel shows it for $p_0=1/4$. The dots denote the optimal demand at $\pi=\EE _0[S]$ for the ambiguity seeker.         
 }
\end{figure}  
Figure \ref{fig:demandfunctions} illustrates that it is optimal for the ambiguity averse (blue dotted line) not to trade at prices below the prior-expected payoff, while both the ambiguity neutral and the ambiguity seeker prefer long positions. Preference for ambiguity makes the optimal position of the ambiguity seeker slightly higher than the one of the ambiguity neutral. At prices above the prior-expected return, such that risk premia defined as the difference between the prior-expected payoff and the price are negative, the ambiguity seeker demands the most negative position, and the ambiguity averse the least negative (albeit different from zero). The figure also illustrates the discontinuity in the demand function under ambiguity when the price equals its $\PP _0$-expectation.

\subsection{Compensation for ambiguity}
In this section, we develop how much of her wealth an investor might be willing to give up to resolve uncertainty, and how this compensation relates to the extant notion of risk.

Fix  price $ \pi$ and trading strategy $ \theta$, not necessarily optimal. From \citet[][]{LeRoyWerner2000}, we can define total compensation $\rho$ for uncertainty and risk as
\begin{equation}\label{E:total premium}
 V_\alpha( \theta)=\EE_0[u\left(W^{ \theta}\right)]-\delta\SS_0[u\left(W^{ \theta}\right)]=:u(w_0-\rho (\delta,  \pi,  \theta)),
\end{equation}  
where we recall that the parameter $\delta$ encodes ambiguity concerns, as $\delta=\sqrt{d}(2\alpha-1)$. Analogously, conventional compensation $\epsilon$ for risk reads
\begin{equation}\label{E:risk premium}
 \EE_0[u\left(W^{ \theta}\right )]=:u(w_0-\epsilon ( \pi,  \theta)).
\end{equation} 
For given $\delta$, trading strategy $ \theta\neq  0$ and prices $ \pi$, the marginal compensation for uncertainty is then naturally defined  as
\begin{equation}\label{eq:defambiguitycomp}
\Delta:=\rho-\epsilon.
\end{equation}
From this definition, we can directly infer  
\begin{equation}
 u(w_0-\epsilon ( \pi,  \theta))-u(w_0-\rho (\delta,  \pi,  \theta))=\delta\SS_0[u\left(W^{ \theta}\right)],
\end{equation} 
 from combining  \eqref{E:total premium} with \eqref{E:risk premium}. 
If $u$ is also assumed increasing, we get that the positivity of $\SS_0$ and the above that compensation for ambiguity is negative for $\alpha < 1/2$ (i.e.~$\delta< 0$), positive when $\alpha > 1/2$ (i.e.~$\delta> 0$) and zero if and only if $\alpha=1/2$ (i.e.~$\delta= 0$).

For example, under exponential utility we have that $\rho (\delta,  \pi,  \theta)= \theta\cdot \pi-\nu(  \theta)$ (recall that $\nu(  \theta)$ is the buyer's certainty equivalent, see \eqref{eq:indifferenceexponentialutil}). In fact, 
\begin{equation}\label{E:ambiguity premium}
\Delta(\delta, \pi, \theta)=\Delta(\alpha, \theta)=\frac{1}{\gamma}\ln\left(1+\delta\frac{\SS_0\left[e^{-\gamma \theta \cdot S}\right]}{\EE_0\left[e^{-\gamma \theta \cdot S}\right]}\right).
\end{equation}
We then get that, under exponential utility, the ambiguity premium \eqref{E:ambiguity premium} has the following interesting connection to the preference associated with the certainty equivalent 
\[\Delta(\alpha,\theta)=-\nu(\theta)-\frac{1}{\gamma}\log \EE_0\left[e^{-\gamma\theta\cdot S}\right ].\]
Note that the second term of the right-hand-side is the certainty equivalent under ambiguity neutrality. In other words, we conclude that the marginal compensation for uncertainty is the difference between the certainty equivalents with and without ambiguity (ambiguity effect on risk measurement is recently studied in \citet{CarPesVand22}, among others). This difference becomes positive for the ambiguity averse and negative for the ambiguity seeker, with its magnitude being an empirical question.

\subsection{A comment on risk-sharing under ambiguity}\label{sec:risk-sharing}

We have already stated that investors' random endowments give motives for them to mutually beneficial trading (with or without ambiguity). The situation is much simpler when investors' endowments are units of the tradeable asset, for instance, in the case for two investors,  $\theta_1,\,\theta_2\,\in\mathbb{R}$.

Exponential utility again allows a high degree of tractability.  In this case,  the equilibrium quantity $\theta^*$ solves the following problem
\[\theta^*=\underset{\theta\in\reals}{\argmax}\{\rho_1(\alpha_1,\theta_1+\theta)-\rho_2(\alpha_2,\theta_2-\theta)\}\]
where $\rho_i$ stands for the (total) compensation of investor $i$ defined in \eqref{E:total premium}. Interestingly enough, we get that when $\delta_1=\delta_2$ and $\PP_{0,1}=\PP_{0,2}$ i.e., investors have the same ambiguity parameter and the initial probability measure, the equilibrium quantities do not depend on $\delta$. Indeed, a calculation shows that  
$$\theta^*=\frac{\gamma_2}{\gamma_1+\gamma_2}\theta_2-\frac{\gamma_1}{\gamma_1+\gamma_2}\theta_1=\frac{\gamma_2}{\gamma_1+\gamma_2}(\theta_1+\theta_2),$$
where $\gamma _i$ are the risk aversion parameters. 

When $\delta_1\neq\delta_2$, this is not the case, and we can conclude that ambiguity reduces the efficiency of risk-sharing,  even when  optimal sharing does not depend on the reference distributions. For an illustrative case assume that the first investor is ambiguity-neutral (i.e., $\delta_1=0$), and the second one is ambiguity-averse, but with no existing exposure (i.e., $\theta_2=0$). The second investor plays the role of an insurer, and without any existing risk there is an interval of prices within which she is not willing to trade $(\upr_2,\,\opr_2)$. On the other hand, if $\theta_1>0$ the reservation price $\pi_1$ (unique due to ambiguity neutrality) of the first investor is less than $\EE_0[S]$, since buying the asset reduces the existing risk. Therefore, there is no trading (no risk transfer), if and only if $\pi_1\geq\upr_2$, which holds for sufficiently large values of $\delta_2$. This implies that ambiguity aversion may shrink the benefits of risk sharing, even when the optimal risk-sharing transaction (under no ambiguity) does not depend on the probability measure(s) of the participating investors.

\bigskip

\section{Market-Clearing Equilibrium under Ambiguity}\label{sec:equilibrium}

In this section, we  study market clearing conditions on a single risky asset, considering  $m\geq 2$ different types of investors (indexed with $i\in \II:=\{1,...,m\}$). Our main goal   is to examine how ambiguity affects an equilibrium among a finite number of price-taking investors, and to analyse whether,  ambiguity seekers need to be present in an economy for non-zero equilibrium transactions. The latter question is motivated from the increased trading propensity of an ambiguity seeker shown in the previous sections. Types of investors may differ with respect to the reference probability measure $\PP^i_{0}$, the divergence parameter $c_i$, ambiguity preference $\alpha_i$ and utility $u_i$. We will concentrate also on the case where  there is heterogeneity only with respect to their ambiguity preference $\alpha _i$.

 Assuming given  supply $\theta _0$ for the tradeable asset, the equilibrium price  $\pi^*$ is defined as the price that clears the market,   
\begin{equation}\label{eq:equilibrium_definition}
\sum _{i\in\II}\tstar_i(\pi^*)+\theta _0=0.
\end{equation}
The majority of the literature  concentrates on an exogenously given positive supply $\theta _0$ of the asset, that can be met also with all participating investors entering long positions. We focus on the case $\theta _0=0$, implying that at least one of the  participating agents must be able to (short) sell the asset in question. It also implies that both, supply and demand, arise endogenously in equilibrium, with  the magnitude $|\tstar_i(\pi^*)|$ of the positions being of particular interest. Derivatives markets, such as options markets accommodate instruments with zero net supply. 

We start with a result given in terms of reservation intervals among ambiguity-averse and ambiguity-neutral investors. The virtue of a statement in terms of reservation intervals lies in their interpretation as \emph{observable} bid ask spreads, rather than abstract notions such as utility functions.  They can however  readily be associated with different traits of the investors participating in the market, such as ambiguity, or risk aversion, according to the results in the current paper.  
\begin{thm}[Market-clearing equilibrium under ambiguity aversion]\label{Thm:equilibrium}
Suppose $\theta _0=0$, and that $\alpha_i\in[1/2,1]$, for each $i\in\II$. There exists an equilibrium price $\pi^*\in(\underline{S},\overline{S})$. The equilibrium positions $\theta _i, \, i\in \mathcal I$, are different from  zero if and only if at least two investors' reservation intervals do not overlap. That is, 
\begin{equation}\label{eq:condition for equilibrium}
\underset{i\in\II}{\min}\,\opr_i<\underset{i\in\II}{\max}\,\upr_i.
\end{equation}
\end{thm}

From FOC \eqref{eq:FOC_measure}, we have that any equilibrium price can be written in terms of an equilibrium pricing measure and equilibrium subjective beliefs. It is easy to check that $\pi^*$ lies outsides of both investors' reservation intervals if and only if we have non-zero trading.  

At this point, it is important to recall that reservation bound prices $\upr$ and $\opr$ are invariant to the utility function and the initial (non-random) wealth. This directly implies that if investors agree on the $\PP _0$ expectation of the tradeable asset, the equilibrium is no trade (as in the case of no-ambiguity concerns). The main difference to the absence of ambiguity is that disagreement about the expected value remains a necessary, but not sufficient condition for mutually beneficial trading. 

As a simple example to describe such a situation, consider the case where investors disagree on the expectation (say $\EE_1[S]< \EE_2[S]$), but agree on variance. Already shutting off ambiguity, with $\alpha_1=\alpha_2=1/2$, there is a non-zero trading equilibrium. However, when $\delta_i\neq 0$ for at least one, such that they are ambiguous about the prevailing probability law,  they will trade if and only if their Sharpe ratios are sufficiently different, and in particular
\begin{equation}\label{ineq:sharpe}
\delta_1+\delta_2<\frac{\EE_{2}[S]}{\SS _0[S]}-\frac{\EE_{1}[S]}{\SS_0[S]}.
\end{equation}
In other words, ambiguity makes investors more hesitant to take  risky positions (their required premium is increased). Condition \eqref{ineq:sharpe} implies that the more pressing the ambiguity concerns, the higher the necessary (and sufficient) difference of Sharpe ratios.

\subsection{Equilibrium among ambiguity-averse and ambiguity-seeking investors}
A preference for ambiguity yields  qualitatively very different demand functions compared to those of ambiguity-averse  (see the indicative Figure \ref{fig:demandfunctions}). In particular, even when the value function $V$ is not locally concave,  demand is discontinuous as zero. This implies that an ambiguity seeker does not have a reservation price, and also that the range of her demand function does not include positions close to zero. Intuitively, an ambiguity seeker is always willing to trade,  and her demand is bounded away from zero. 

\begin{figure}
 \input{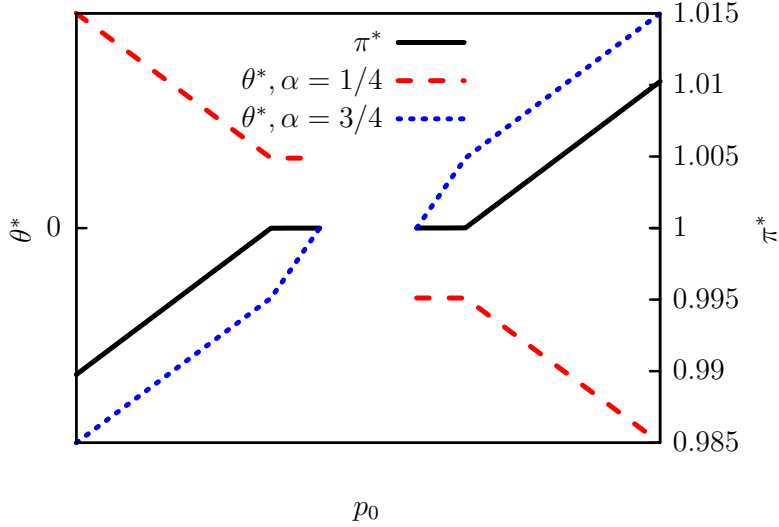}
  \caption{\label{fig:equilibrium}Equilibrium in binomial model. The figure shows equilibrium price $\pi^*$ in a two-agent economy with state space $S_g=1.1, S_b=0.9$ with $\alpha _1=1/4$ and $\PP_{1}(S_g)=1/2$, and $\alpha _2 = 3/4$ and $\PP_{2}(S_g)=p_0$. Both market participants employ log utility. The equilibrium optimal allocations are depicted on the left y-axis, while the equilibrium price is depicted on the right y-axis.}   
\end{figure}
Figure \ref{fig:equilibrium} illustrates how a market-clearing equilibrium can arise from differences in reference probabilities and ambiguity preferences. It shows equilibrium prices and positions in a binomial economy with states $S_g=1.1$ and $S_b=0.9$ for two investors as a function of $p_0$, where $\PP_{0,1}(S_g)=1/2, \alpha _1=1/4$ and $\PP_{0,1}(S_g)=p_0, \alpha _2=3/4$, and the initial supply $\theta _0=0$.  In line with the theory developed in this section, the two investors' prior beliefs are sufficiently diverse to accommodate a first-best equilibrium that is optimal for both counterparties. The magnitudes of the positions are increasing in the difference of the beliefs, and when $\PP_{1}=\PP _{2}=1/2$, there is no trade. The equilibrium prices take an option-type shape as a function of the reference probability assigned to $S_g$ by the ambiguity averse, that reflects which prices the ambiguity seeker is willing to accept in order to satisfy her desire to trade in long \emph{or} short positions despite her fixed reference probability measure.

While preference for ambiguity increases  the desired trading volume, it does not necessarily make  non-zero equilibrium trading more likely. This mainly stems from the aforementioned jump of the demand around the expectation of the tradeable asset under the reference probability measure $\PP _0$. In particular, the inequality $\opr_i<\opr_j$ between an ambiguity seeker and an ambiguity averse does not necessarily imply the existence of an equilibrium price. This means that condition \eqref{eq:condition for equilibrium} is not sufficient for  market clearing when the heterogeneity in ambiguity preferences among investors is large enough to accommodate both types. We substantiate this argument in the following counter-example. 

\begin{exa}
Consider a binomial economy with $S_g>S_b$  representing the outcomes of the risky asset, and two investors with exponential utility, but with different risk aversion coefficients $\gamma_1\neq\gamma_2$. Assume that the first investor is an ambiguity seeker with $\delta_1<0$, and the second is ambiguity averse with $\delta_2>0$. Investors have sufficiently distinct reference probability measures, such that \eqref{eq:condition for equilibrium} holds, and in particular we assume that $\opr_2<\upr_1$, where 
\begin{align*}
\opr_i &=p_iS_g+(1-p_i)S_b+|\delta_i|(S_g-S_b)\sqrt{p_i(1-p_i)},\\
\upr_i &=p_iS_g+(1-p_i)S_b-|\delta_i|(S_g-S_b)\sqrt{p_i(1-p_i)},
\end{align*}
and $p_i$ stands for the reference probability investor $i$ assigns to $S_g$. Based on Proposition \ref{P:demand_function} and the discussion in Section \ref{sub:binomial}, we get that  potential equilibrium prices $\pi$ are  within the interval $(\opr_2,\EE_1[S])$, where $\theta^*_1(\pi)>0$ and $\theta^*_2(\pi)<0$. Simple calculations yield that 
\begin{align*}
\theta^*_1(\pi)&=\frac{1}{\gamma_1(S_g-S_b)}\ln\left(\frac{(S_g-\pi)(p_1+|\delta_1|\sqrt{p_1(1-p_1)})}{(\pi-S_b)(1-p_1-|\delta_1|\sqrt{p_1(1-p_1)})}\right), \text{ and }\\
\theta^*_2(\pi)&=\frac{1}{\gamma_2(S_g-S_b)}\ln\left(\frac{(S_g-\pi)(p_2-|\delta_2|\sqrt{p_2(1-p_2)})}{(\pi-S_b)(1-p_2+|\delta_2|\sqrt{p_2(1-p_2)})}\right).
\end{align*}
Note that $\theta_2^*(\pi)$ is strictly decreasing for prices higher than $\opr_2$, while $\theta_1^*(\pi)$ is strictly decreasing for prices less than $\EE_1[S]$. This implies that the clearing condition $\theta_1^*(\pi)=-\theta_2^*(\pi)$ holds if and only if $-\theta_2^*(\EE_1[S])\geq\theta_1^*(\EE_1[S])$, when the ranges of the functions $\theta_1^*(\pi)$ and $-\theta_2^*(\pi)$ for prices  $\pi\in(\opr_2,\EE_1[S])$ overlap. 

Consequently, we need to verify that $-\theta_2^*(\EE_1[S])<\theta_1^*(\EE_1[S])$ is attainable under the condition $\opr_2<\upr_1$ in order to complete the counterexample. To facilitate this step, we may simplify the calculations further, by assuming that $S_g=1$, $S_b=0$ and that $|\delta_i|(S_g-S_b)\sqrt{p_i(1-p_i)}=\mu$, for both $i=1,2$. The desired inequality becomes,  
\begin{equation}\label{eq:counterexample1}
 \frac{1}{\gamma_2}\ln\left(\frac{p_1(1-p_2+\mu)}{(1-p_1)(p_2-\mu)}\right)<\frac{1}{\gamma_1}\ln\left(\frac{(1-p_1)(p_1+\mu)}{p_1(1-p_1-\mu)}\right),
\end{equation}
under the condition $p_2+\mu<p_1-\mu$. Finally,  by imposing $p_1=1-p_2$, and common $|\delta_i|=\delta$, the inequality $p_2+\mu<p_1-\mu$ becomes $0<p_2^2-p_2+(4(1+\delta^2))^{-1}$, which holds for all $p_2\in(0,\frac{1}{2}(1-\delta/\sqrt{1+\delta^2}))$. Under this specification and a choice $p_2$ within this interval, \eqref{eq:counterexample1} holds when $1/\gamma_1$ is sufficiently large. In other words, even if the initial probability measures of the investors are very different,
there is no equilibrium transaction between them if the ambiguity seeker is also sufficiently risk tolerant. In this case, her demand at the discontinuity point (that is, when $\pi=\EE_0[S]$) is larger than the corresponding supply of the ambiguity-averse investor. 
\end{exa}

The above counterexample is a first indication about the mechanics of equilibrium trading among ambiguity-seeking, and ambiguity-averse investors. In particular, we readily get from Propositions \ref{P:demand_function} and \ref{P:seeker} that a non-zero equilibrium transaction between an ambiguity seeker (say investor 1) and an ambiguity averse (say investor 2) exists if, in addition to $\opr_2<\upr_1$, it holds that $\theta_2^*(\EE_1[S])<-\theta_1^*(\EE_1[S])$. The latter means that the discontinuity point of the ambiguity seeker's demand is sufficiently low to intersect with her ambiguity-averse counterparty's supply.

\subsection{Preference for ambiguity as a sole source for trading}\label{sub:second-best equilibrium}
Heterogeneity in reference beliefs is an important component in the  equilibria discussed in the previous sections, and as illustrated in Figure \ref{fig:equilibrium}. It is natural to ask, whether  equilibria exist also if prior reference beliefs $\PP_0$ are not sufficiently diverse, such that reservation intervals overlap. If additionally $\theta_0=0$,  investors desire the same type of positions (long or short), independently of their ambiguity preference, yielding a no-trade equilibrium. However, as we have seen in Propositions \ref{P:seeker} and \ref{p:binomial_seeker}, ambiguity seekers are willing to take short (resp.~long) position on the asset, even if the price is lower (resp.~higher) than $\EE_0[S]$, in the sense that  their value functions are higher than the initial status quo for positions at both sides of trading. This comes in sharp contrast with the ambiguity averse/neutral investors, where any positive (resp.~negative) position on the risky asset at price higher (resp.~lower) than $\EE_0[S]$ deteriorates their value function. 

Motivated from these observations, we infer that there is room for mutually beneficial Pareto-optimal trading, even if investors have the exact same characteristics  except for their preference for ambiguity encoded in the coefficient $\delta$.  
\begin{defn}[Pareto-optimal second-best equilibrium]\label{def:Pareto}
 A price $\tilde {\pi}$ is called a second-best Pareto-optimal equilibrium price, if
\[\sum _{i\in\II} \tilde {\theta}_i(\tilde \pi)+\theta _0=0,\]
where $\tilde {\theta}_i(\pi):=\theta_i^*( \pi)$, for $\alpha _i \geq 1/2$ and  
\begin{equation}
\text{for }\alpha_i<1/2,\qquad
\tilde {\theta}_i(\pi):=\begin{cases}
                                \underset{\theta\leq 0}{\argmax}\, V_{\alpha _i}(\theta), & \EE _i[S] \geq  \pi,  \\
                                \underset{\theta\geq 0}{\argmax}\, V_{\alpha _i}(\theta), & \EE _i[S] <  \pi.
                               \end{cases}
\end{equation}
\end{defn}

The  equilibrium according to Definition \ref{def:Pareto} is second-best in the sense that the  ambiguity seeker is restricted to one side of the trade: long, if the price is higher than the initial expectation,  and short, if the price is lower than the initial expectation. 
Such an  equilibrium is reasonable, when ambiguity seekers are involved in the transaction, due to the fact that these investors are always better off trading on either side of the trade, if prices are not very high (higher than $\opr_i$), or very low (lower than $\upr_i$), than not trading at all. Furthermore, it follows directly from the  definition of the second-best Pareto equilibrium that there always exists a non-zero equilibrium between two ambiguity seekers. The restriction of the portfolio weight  can be used to identify institutional constraints as mechanisms to induce an equilibrium. For instance, the investment strategy of a pension or a mutual fund is usually restricted to long positions. 

\begin{figure}
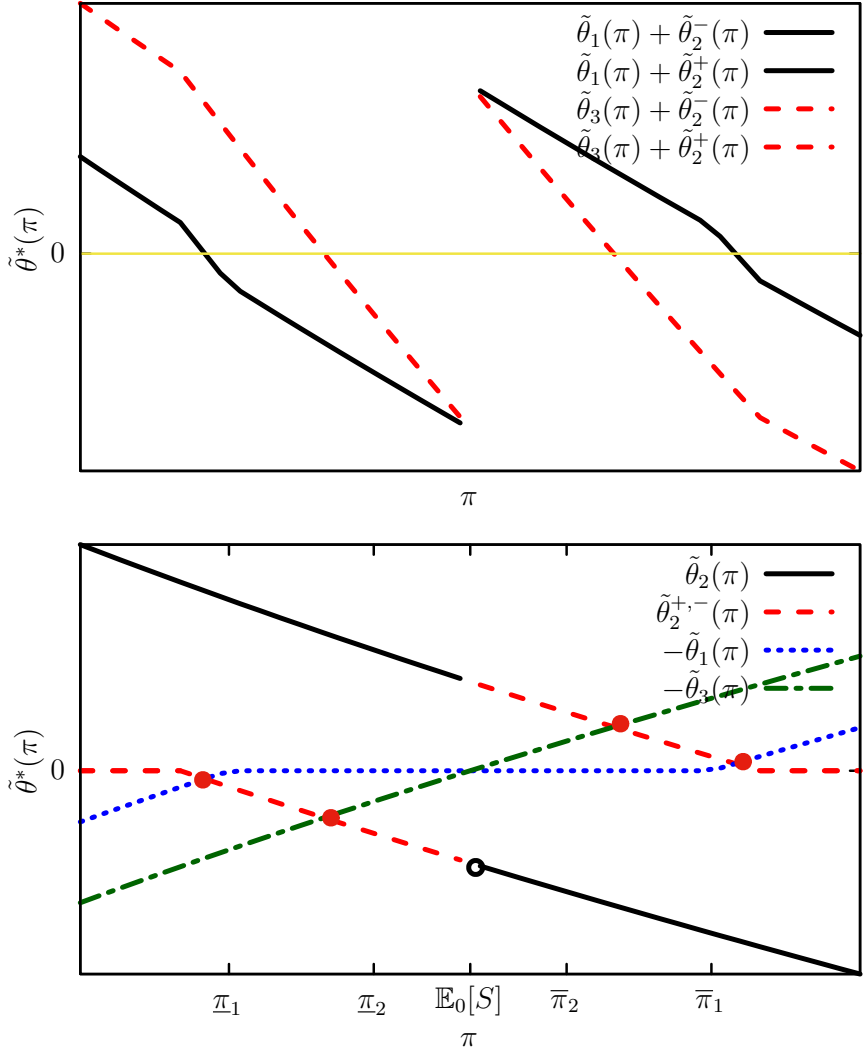

  \scalebox{1}{\input{demand_sb_lo}}    
  \scalebox{1}{\input{demand_fb_sb}}             
   \caption{\label{fig:secondbestnetdemand}Second-best net demand. For this figure, three agents are considered in a binomial economy:  ambiguity-averse ($\alpha_1=3/4$), ambiguity-seeking ($\alpha_2=2/5$), and ambiguity-neutral ($\alpha_3=1/2$), while all three agree on their reference prior $\PP _0(S_g)=p_0=1/4$, initial wealth $1$ and power utility with parameter $\gamma=2$. The asset in question takes the values $S_g=1.1$ and $S_b=0.9$.  The top figure shows the net supply of a binomial asset, where agent one and agent three choose their first-best optimum, and agent two chooses her second-best. The bottom panel shows the corresponding positions. The red dots mark the equilibrium allocations according to Definition \ref{def:Pareto}. The black dot marks the optimal allocation at $\pi=\EE_0[S]$ of the ambiguity seeker.
   }         
 \end{figure}

Figure \ref{fig:secondbestnetdemand} illustrates a second-best Pareto-optimal equilibrium according to Definition \ref{def:Pareto} in the binomial economy introduced in Section \ref{sub:binomial} for three   investors participating in the market, one ambiguity-averse, one ambiguity-neutral, and one ambiguity-seeking. Specializing to two market places  with two investors each,   the figure shows the existence of two second-best equilibrium prices in these economies. The first two are between the ambiguity seeker and the ambiguity-neutral. The second two are between the ambiguity seeker and the ambiguity-averse. A non-zero second-best Pareto-optimal equilibrium always exists among two ambiguity seekers, and therefore not considered as an example. 
Ambiguity aversion controls the size of the position, with the intuitive relation that less ambiguity aversion yields higher positions.

The next result  is intended to be a follow-up to Proposition \ref{p:binomial_seeker}. It deals with the second-best equilibrium under a binomial model when there exists a single ambiguity seeker. 
\begin{thm}\label{P:Second-best_binomial}
Let $|D|=2$, $K=1$, $\theta_0=0$ and assume that $\alpha_1<1/2$ and $\alpha_i\geq 1/2$, for all $i\in\II$, $i>1$. A sufficient condition for the existence of a non-zero second-best Pareto optimal equilibrium price $\tilde{\pi}\in(\underline{S},\overline{S})$ is 
\begin{equation}\label{eq:condition for second best}
\opr_1>\underset{i\in\II\setminus\{1\}}{\max}\,\opr_i\qquad\text{or}\qquad \upr_1<\underset{i\in\II\setminus\{1\}}{\min}\,\upr_i.
\end{equation}
\end{thm}

\smallskip 

The simple binomial setting is ideal to gain intuition about possible trading  equilibria under ambiguity. 
Assume that there are two investors with common reference distribution $\PP_0$ and common divergence parameter $c$, common utility function $u$, but different ambiguity aversion $a_1<1/2$ and $a_2>1/2$. It is clear from their demand functions that for prices $\pi >\EE_0[S]$ (resp.~$\pi<\EE_0[S]$), they both want to sell (resp.~buy). Hence, the first-best equilibrium definition, i.e., $\tstar_1(\pi^*)+\tstar_2(\pi^*)=0$, has no solution. 

However, we can describe second-best equilibria that make both investors better off than with no trading. For this, we just need the condition $\alpha_1+\alpha_2<1$. Indeed, the latter is equivalent to the inequality 
\[\opr_2=\EE_0[S]+\delta_2\SS_0[S]<\opr_1=\EE_0[S]-\delta_1\SS_0[S].\] 
This means that for prices $\pi\in(\opr_2,\opr_1)$,  the ambiguity-averse is willing to sell, while the ambiguity seeker will increase her objective if she buys (a second-best Pareto-optimal equilibrium). Similarly, we have that 
$\alpha_1+\alpha_2<1$ implies $\upr_1<\upr_2$ and there is a second second-best equilibrium where the  ambiguity-averse buys, and the ambiguity seeker sells. We state this observation as a corollary.

\begin{cor}\label{c:binomial}
Depart from the assumptions made in Theorem \ref{P:Second-best_binomial}. If we further assume that all investors have the same initial probability measure $\PP_0$, divergence parameter $c$, utility function $u$ and initial wealth $w_0$, a sufficient condition for the existence of a non-zero second-best Pareto-optimal equilibrium price $\tilde{\pi}\in(\underline{S},\overline{S})$ is 
\begin{equation}\label{eq:condition for second best common beliefs}
\alpha_1+\underset{i\in\II\setminus\{1\}}{\min}\,\alpha_i<1.
\end{equation}
\end{cor}

We should emphasize that the non-zero second-best equilibrium exists without assuming anything in particular about investors' utility functions, since the sufficient conditions depend solely on the prices $\upr_i$ and $\opr_i$. 

As we have seen, the binomial model guarantees that the objective function of an ambiguity seeker is always concave for positive and negative positions on the risky asset (as long as the utility is strictly concave). This local concavity of $V_\alpha$, together with condition \eqref{eq:condition for second best common beliefs}, yields that sufficient divergence of ambiguity is a source of mutually beneficial trading among traders who are all else homogeneous. 

In general, $V_\alpha$ is not concave for $\alpha\in(0,1/2)$ (see the related discussion in  Section \ref{sec:ambiguity seeking}). However, the fact that sufficient heterogeneity in ambiguity preferences is a source for mutually beneficial trading among homogeneous traders still holds, but in local terms. In other words, condition \eqref{eq:condition for second best common beliefs} always gives a non-zero second-best Pareto-optimal equilibrium, when the set of admissible positions on the tradeable asset is sufficiently restricted. For this, we need to slightly extend the definition of second-best equilibrium.
\begin{defn}[Local Pareto-optimal second-best equilibrium]\label{D:Local_equilibrium}
A price $\tilde{\pi}_{loc}$ is called a \textit{local} second-best Pareto-optimal equilibrium price, if there exists a subset $\Theta_r\subset\Theta$ such that 
\[\sum _{i\in\II} \theta^r_i(\tilde{\pi}_{loc})+\theta _0=0,\]
where $\theta^r_i(\pi)=\underset{\theta\in\Theta_r}{\argmax}\,V_{\alpha_i}(\theta),$ for $\alpha _i \geq 1/2$ and  
\begin{equation*}
\text{for }\alpha_i<1/2,\qquad
\theta^r_i(\pi):=\begin{cases}
                                \underset{\theta\in\Theta_r\cap(\underline{\Theta},0]}{\argmax}\, V_{\alpha _i}(\theta), & \EE _i[S] \geq  \pi,  \\
                                \underset{\theta\in\Theta_r\cap[0,\overline{\Theta})}{\argmax}\, V_{\alpha _i}(\theta), & \EE _i[S] < \pi.
                               \end{cases}
\end{equation*}
\end{defn}
From Definition \ref{D:Local_equilibrium}, we can directly state the following
\begin{prop}\label{P:Local_Second-best}
Let $K=1$, $\theta_0=0$, $\alpha_1<1/2$ and $\alpha_i\geq 1/2$, for all $i\in\II$, $i>1$ and impose condition \eqref{eq:condition for second best}. Then, all prices $\pi$ in $(\max_{i\in\II\setminus\{1\}}\,\opr_i,\opr_1)$ or in $(\upr_1,\min_{i\in\II\setminus\{1\}}\,\upr_i)$, produce non-zero local second-best Pareto optimal equilibria. In particular, \eqref{eq:condition for second best common beliefs} is sufficient for the existence of a non-zero local second-best Pareto optimal equilibrium for homogeneous investors. 
\end{prop}

We emphasize that the existence of local second-best equilibria holds for arbitrarily many states $D$, and for  arbitrary strictly concave utility functions, even when the ambiguity seeker's value function is not locally or globally concave. In this quite general setting, when the price is higher than the common expectation $\EE_0[S]$, but not higher than $\opr_1$, the ambiguity seeker's value function is increasing for positive $\theta$. Continuity of demand function on positive positions (deduced directly from FOC \eqref{eq:FOC}), yields that the ambiguity seeker always has a non-zero locally optimal demand, which is the source of this non-zero mutually beneficial transaction. 

The shape of $\Theta_r$ in order to establish a market-clearing equilibrium motivates a type of inverse problem, that asks: What are the market frictions and institutional restrictions necessary to explain observed prices and trading volume together? These could range from shortselling and  trading volume restrictions, for instance from optimal execution algorithms, to risk management constraints. 

%



\bigskip

\section{Conclusion}\label{sec:conclusion}
We investigate  optimal portfolio allocations under  $\alpha$-MEU preferences,  where agents' nonparametric beliefs obtain in the neighbourhood of a reference distribution in dependence of the trading strategy chosen. In this setting, market participants distinguish between preferences for risk and uncertainty/ambiguity. The $\alpha$-MEU framework further allows the distinction between ambiguity-seeking $(\alpha<1/2)$, and ambiguity-averse ($\alpha>1/2$) ambiguity preferences. 

We derive the value function under $\alpha$-MEU in closed form as a superposition of expected utility and an additional term that is proportional to the standard deviation of utility of terminal wealth. We show that this value function is strictly concave under ambiguity aversion, and allows rapid and globally optimal portfolio allocation rules also in the multi-asset case. In the case of a preference for ambiguity, for $\alpha >1/2$, we provide an algorithm that is guaranteed to converge to  locally optimal portfolio rules. For $\alpha=1/2$, $\alpha$-MEU preferences reduce to standard expected utility in our setting, justifying the term \emph{ambiguity neutrality}. We use this natural benchmark to motivate a definition of compensation for ambiguity that is separate from the compensation for risk.

From the tractability of our framework, we investigate the existence of market-clearing equilibria, in particular for zero net supply settings, such as option markets. Such equilibria  arise foremost through differences of beliefs. When beliefs are close, or even identical, for instance when reference distributions can be estimated with high precision, we prove existence of another,  Pareto-optimal equilibrium, provided at least one ambiguity-seeking individual participates in the market place. These Pareto equilibria crucially depend on restrictions on the set of trading strategies, such as short-selling or risk management constraints. They  motivate a type of inverse problem, where the market frictions are chosen as parameters to yield equilibria compatible with observed prices. 

Our analysis provides the basis for an array of subsequent future research, both methodological and empirical. Methodologically, sufficient conditions for global concavity of value functions for ambiguity seekers are the most pressing. 
Empirical  investigations of secondary markets with zero net supply offer themselves, such as options markets, to build upon recent studies in asset markets, such as \citet{brennerizhakian18, augustinizhakain19}, in the presence of market frictions.  

\bigskip

\begin{appendix}
 \section{Proofs}
 \subsection{Proposition \ref{P:optimization}}
 \begin{proof}
We start with the Lagrangian of program \eqref{eq:min_simple}, 
\begin{equation*}
 L(p_1,\ldots, p_n, \nu,\eta):=\sum _{s=1}^{n}u(W_s) p_s
 -\nu \brac{\sum _{s=1}^{n}p_s-1}+\eta \brac{\sum _{s=1}^{n}\frac{p^2_s}{p_{0s}} -c},
\end{equation*}
with the corresponding Karush-Kuhn-Tucker (KKT) conditions 
\begin{equation}\label{KKT conditions}
\begin{split}
 L_{p_s}& \geq 0,\quad p_s\geq 0\quad\text{and}\quad p_sL_{p_s}=0\\ 
 L_{\eta} &\leq 0,\quad\eta\geq 0\quad\text{and}\quad \eta L_{\eta}=0\\ 
L_{\nu}&= 0.
\end{split}
\end{equation}
Below, we show that under Assumption \ref{A:condition for c}, the solution of the KKT satisfies $L_{p_s}=0$ and $L_{\eta}=0$ (which means that the divergence constraint is binding). We calculate
\[L_{p_s}= u(W_s)-\nu+2\eta\frac{p_s}{p_{0s}}=0,\]
which implies that $p_s=p_{0s}(\nu-u(W_s))/2\eta$.
The condition $L_{\nu}= 0$ is equivalent to $\sum _{s\in D}p_s=1$, and together with $L_{\eta}=0$ the Lagrangians are 
\begin{equation}\label{eq:eta,nu_min}
 \underline{\nu}=\frac{\SS_0[u(W)]}{\sqrt{d}}+\EE_0[u(W)], \quad\text{ and }\quad\underline{\eta}=\frac{\SS_0[u(W)]}{2\sqrt{d}},
\end{equation}
This gives the unique solution of KKT if $p_s=p_{0s}(\nu-u(W_s))/2\eta\geq 0$, for each $s\in D$. This is indeed the case, since Lemma \ref{L:condition} guarantees that 
\begin{equation}\label{eq:constrain for max_u}
\max\{u(W_{s})\}<\EE_0[u(W)]+\frac{\SS_0[u(W)]}{\sqrt{d}}.
\end{equation}
Then, simple calculations imply \eqref{eq:p_min}. 

On the other hand, the Lagrangian of the maximization problem \eqref{eq:maximization} is 
\begin{equation*}
 L(p_1,\ldots, p_n, \nu,\eta):=\sum _{s=1}^{n}u(W_s) p_s
 -\nu \brac{\sum _{s=1}^{n}p_s-1}-\eta \brac{\sum _{s=1}^{n}\frac{p^2_s}{p_{0s}} -c},
\end{equation*}
and the KKT conditions are 
\begin{align*}\label{KKT conditions_max}
L_{p_i}&\leq 0,\quad p_s\geq 0\quad\text{and}\quad p_sL_{p_s}=0\\ 
L_{\eta}&\geq 0,\quad\eta\geq 0\quad\text{and}\quad \eta L_{\eta}=0\\ 
L_{\nu}&= 0.
\end{align*}
Again, we  have that Assumption \ref{A:condition for c} yields that the unique solution satisfies $L_{p_s}= 0$ for each $s\in D$ and $L_{\eta}= 0$. Indeed, these conditions, together with $L_{\nu}= 0$, give that 
\begin{equation}\label{eq:eta,nu_max}
 \overline{\nu}=\EE_0[u(W)]-\frac{\SS_0[u(W)]}{\sqrt{d}}, \quad\text{ and }\quad\overline{\eta}=\frac{\SS_0[u(W)]}{2\sqrt{d}},
\end{equation}
and $\overline{p}_s$ is given by \eqref{eq:p_max}, which is non-negative if
\begin{equation}\label{eq:constrain for min_u}
\min\{u(W_{s})\}>\EE_0[u(W)]-\frac{\SS_0[u(W)]}{\sqrt{d}},
\end{equation}
holds true. However, the latter is proven to hold under Assumption \ref{A:condition for c} in Lemma \ref{L:condition}.

Having proved \eqref{eq:p_min} and \eqref{eq:p_max}, it is then matter of calculations to get that
\[V_\alpha(W)=\alpha \sum_{s=1}^n u(W_s)\underline{p}_s+(1-\alpha)\sum_{i=1}^nu(W_s)\overline{p}_s\]
equals  \eqref{eq:V_a nice}.
\end{proof}
\subsection{Lemma \ref{L:condition}}
\begin{proof}
We use the simplified notation $u(W_{s}):=u_s$, for each $s\in D$ and $u:=u(W)$, and without loss of generality assume that $u_1=\max_{s\in D}u_s$. For the first part we need to show that
\begin{align}\label{ineq1}
\nonumber u_1-\EE_0[u]&<\frac{\SS_0[u]}{\sqrt{c-1}}, \text{ that is }\\
(c-1)(u_1-\EE_0[u])^2&<\VV_0[u].
\end{align}
For these inequalities to hold, a sufficient condition is $c-1<1/(1-p_{01})-1$, which follows directly from Assumption \eqref{A:condition for c}. Indeed, $c-1<1/(1-p_{01})-1=p_{01}/(1-p_{01})$, hence it is enough to show that
\begin{equation}\label{ineq2}
\frac{p_{01}}{1-p_{01}}(u_1-\EE_0[u])^2\leq \VV_0[u].
\end{equation}
On the other hand, we calculate 
\begin{align*}
\VV_0[u]=\sum_{s\in D}p_{0s}(u_s-\EE_0[u])^2=p_{01}(u_1-\EE_0[u])^2+\sum_{s\geq 2}p_{0s}(u_s-\EE_0[u])^2,
\end{align*}
and \eqref{ineq2} is equivalent to
\begin{equation}\label{ineq3}
\frac{p_{01}^2}{1-p_{01}}(u_1-\EE_0[u])^2 \leq \sum_{s\geq 2}p_{0s}(u_s-\EE_0[u])^2. 
\end{equation} 
We now decompose the LHS as 
\begin{align*}
\frac{p_{01}^2}{1-p_{01}}(u_1-\EE_0[u])^2 &= \frac{p_{01}^2}{1-p_{01}}\left(u_1(1-p_{01})-\sum_{i\geq 2}p_{0i}u_i\right)^2\\
 &=  p_{01}^2(1-p_{01})\left(u_1^2+\tilde{\EE}^2[u]-2u_1\tilde{\EE}[u] \right),
\end{align*}
where $\tilde{\EE}$ denotes the expectation under the measure $\tilde{\PP}:=(0,p_{02}/(1-p_{01}),p_{03}/(1-p_{01}),...,p_{0n}/(1-p_{01}))$ (which is the expectation conditional on the event $u<u_1$). 
Similarly, we decompose the RHS of \eqref{ineq3} as
\begin{align*}
\sum_{s\geq 2}p_{0s}(u_s-\EE_0[u])^2&=\sum_{s\geq 2}p_{0s}\left(\left(u_s-\sum_{i\geq 2}p_{0i}u_i\right)-p_{01}u_1\right)^2\\
 &= \sum_{s\geq 2}p_{0s}\left[\left(u_s-\sum_{i\geq 2}p_{0i}u_i\right)^2+p^2_{01}u^2_1-2\left(u_s-\sum_{i\geq 2}p_{0i}u_i\right)p_{01}u_1\right]\\
&=   \sum_{s\geq 2}p_{0s}\left(u_s-\sum_{i\geq 2}p_{0i}u_i\right)^2+(1-p_{01})p^2_{01}u^2_1-2p^2_{01}u_1\sum_{i\geq 2}p_{0i}u_i\\
&= (1-p_{0i})\left(\sum_{s\geq 2}\frac{p_{0s}}{1-p_{01}}\left(u_s-\sum_{i\geq 2}p_{0i}u_i\right)^2 +p^2_{01}u_1^2-2p^2_{01}u_1\tilde{\EE}[u]\right).
\end{align*}
Therefore, \eqref{ineq3} is equivalent to 
\begin{align*}
p_{01}^2\tilde{\EE}^2[u] &\leq \sum_{s\geq 2}\frac{p_{0s}}{1-p_{01}}\left(u_s-\sum_{i\geq 2}p_{0i}u_i\right)^2\\
p_{01}^2\tilde{\EE}^2[u] &\leq  \tilde{\EE}[u^2]+ \sum_{i\geq 2}p_{0i}u_i -2\sum_{i\geq 2}p_{0i}u_i\tilde{\EE}[u]\\
p_{01}^2\tilde{\EE}^2[u] &\leq  \tilde{\EE}[u^2]+ (1-p_{01})^2\tilde{\EE}^2[u] -2(1-p_{01})\tilde{\EE}^2[u]\\
p_{01}^2\tilde{\EE}^2[u] &\leq  \tilde{\EE}[u^2]-(1-p^2_{01}) \tilde{\EE}^2[u]\\
 0 &\leq  \tilde{\EE}[u^2]-\tilde{\EE}^2[u]  ,
\end{align*}
which holds true. 

Similarly for the second part, we need to show that 
\begin{align*}
u_n-\EE_0[u]&>-\frac{\SS_0[u]}{\sqrt{c-1}}\\
(c-1)(\EE_0[u]-u_n)^2&<\VV_0[u].
\end{align*}
The sufficient condition is $c<1/(1-p_{0n})$, which follows directly from Assumption \eqref{A:condition for c}. Indeed, $c-1<1/(1-p_{0n})-1=p_{0n}/(1-p_{0n})$, hence it is enough to show that 
\[\frac{p_{0n}}{1-p_{0n}}(\EE_0[u]-u_n)^2\leq \VV_0[u].\]
For the above, we just follow a similar decomposition as in the first part of the statement. 
\end{proof}
\subsection{Proposition \ref{P:increasing}}
\begin{proof}
Since $W_1$ stochastically dominates random $W_2$ under probability measure $\PP_0$, there exists a state $s'\in D$ such that $W_{2s'}<W_{1s'}$ and due to the monotonicity of $u$, it holds that $u(W_{2s'})<u(W_{1s'})$. This implies that $\EE_{\PP}[u(W_{2})]<\EE_{\PP}[u(W_{1})]$ for all probability measures $\PP\sim\PP_0$. Recall from Proposition \ref{P:optimization}, that min and max probability measures that solve \eqref{eq:V_a} are equivalent to $\PP_0$. This implies in particular that 
\[\max_{P\in \XX}\EE_P[u(W_{1})]=\EE_{\overline{P}_1}[u(W_{1})]<\EE_{\overline{P}_1}[u(W_{2})]\leq \max_{P\in \XX}\EE_P[u(W_{2})].\]
Similarly, we have that 
\[\min_{P\in \XX}\EE_P[u(W_{2})]=\EE_{\underline{P}_2}[u(W_{2})]>\EE_{\underline{P}_2}[u(W_{1})]\geq \min_{P\in \XX}\EE_P[u(W_{1})],\]
which completes the proof.
\end{proof}
\subsection{Proposition \ref{P:concavity}}
\begin{proof}
Based on representation \eqref{eq:V_a nice}, we first define the function $\MM:\mathbb{R}^D\mapsto\mathbb{R}$ as
\[\MM(X)=\EE_0[X]-\delta\SS_0[X],\]
where in fact $\delta:=\sqrt{d}(2\alpha-1)\geq 0$ as in the main text. For this auxiliary function we have the following result.
\begin{lem}\label{lem:concavity}
The following statements hold 
\begin{itemize}
\item[i.] Function $\MM$ is strictly monotone, in the sense that $\MM(X_2)-\MM(X_1)\geq 0$, if $X_2\geq X_1$ and $\MM(X_2)-\MM(X_1)> 0$, if $X_2\geq X_1$ and for some $D'\subset D$, it holds that $X_2(s)> X_1(s)$ for each $s\in D'$, and  
\item[ii.] for $\delta>0$, $\MM$ is concave on any convex set $\CC\subset\reals^D$ that does not contain zero. 
\end{itemize}
\end{lem}
\begin{proof}
The monotonicity follows directly from \eqref{eq:V_a}, while for the strict monotonicity it is enough to observe from Proposition \ref{P:optimization} that both measures $\underline{P}$ and $\overline{P}$ are equivalent to $\PP_0$. 

For the second item, for any $X,Y\in\mathbb{R}^D$, we define the function $m:\mathbb{R}\mapsto\mathbb{R}$ as $m(t):=\MM(X+tY)$. The second derivative $m''(t)$  is the second derivative of $\MM$ at $X$ in the direction of $Y$, i.e.,
\[m''(t)=\frac{d^2}{dt^2}\left.\MM(X+tY)\right.\]
Assuming that $X\neq 0$ implies that $\SS_0[X]>0$ and hence the aforementioned second derivative exists. In particular,
\[m''(t)=\sum_{i}^D\MM_{ii}(X+tY)Y_i^2+2\sum_{i\neq j}\MM_{ij}(X+tY)Y_iY_j,\]
where $\MM_{ii}$ stands for the second partial derivative of $\MM$ at each $i$th argument. The latter implies that for each $Y\in\mathbb{R}^D$
\[m''(0)=Y'\mathbb{H}_M(X)Y=\delta\frac{\Cov^2(X,Y)-\SS^2_0[X]\SS^2_0[Y]}{(\SS_0[X])^{3}}=\frac{\delta\SS^2_0[Y]}{\SS_0[X]}(\rho^2(X,Y)-1)\leq 0,\]
where $\mathbb{H}_M(X)$ is the Hessian matrix of $\MM$ at $X$ and $\rho$ denotes the correlation coefficient. The latter implies that the Hessian matrix of $\MM$ is negative semi-definite at each $X\in\CC$  and hence that $\MM$ is concave. 
\end{proof}

Returning to the proof of Proposition \ref{P:concavity}, we point out that function $\Theta\ni \theta\mapsto U(\theta):=u(x_0+\theta\cdot(S-p))\in\RR^D$ is pointwise concave for any $\theta\neq 0$ (recall that we assume that  tradeable assets are not linearly dependent). In other words, for each $\theta_1,\theta_2\neq 0 $ and $\lambda\in [0,1]$, it holds that 
\[U(\lambda\theta_1+(1-\lambda)\theta_2)\geq \lambda U(\theta_1)+(1-\lambda)U(\theta_2).\]
Note that, for each $\theta\neq 0$, there is a neighbourhood $\mathcal{N}$ around $\theta$ such that all random variables $\{U(\theta)\,:\, \theta\in\mathcal{N}\}\subset\mathbb{R}^D$ belong to a convex set $\CC$ which does not contain any constants (note that $\SS_0$ is a continuous function). This implies that function $\MM$ is concave over $\CC$ and we use Lemma \ref{lem:concavity} for
\begin{align*}
V_\alpha(\lambda\theta_1+(1-\lambda)\theta_2) &=\MM(U(\lambda\theta_1+(1-\lambda)\theta_2))\geq\MM(\lambda U(\theta_1)+(1-\lambda)U(\theta_2))  \\
&\geq \lambda\MM(U(\theta_1))+(1-\lambda)\MM(U(\theta_2))=\lambda V_\alpha(\theta_1)+(1-\lambda)V_\alpha(\theta_2).
\end{align*}
Note also that the first inequality is strict if also $\theta_1\neq\theta_2$ and $\lambda\in(0,1)$, which shows that $V$ is indeed strictly concave. 

\end{proof}

%

\subsection{Lemma \ref{lem:large theta}}
\begin{proof}
We readily get from \eqref{eq:V_a nice} that for all $\theta\neq 0$ and $\pi\in(\upr,\opr) $ it holds
\begin{align*}
V'_\alpha(\theta)&=\EE_0\left[ u'(W_1)(S-\pi)\left(1-\delta\frac{u(W_1)-\EE_0[u(W_1)]}{\SS_0[u(W_1)]}\right) \right]\\
&=\sum_{S_k\geq \pi}P_{0k}u'(W_{1k})(S_k-\pi)\left(1-\delta\frac{u(W_{1k})-\EE_0[u(W_1)]}{\SS_0[u(W_1)]}\right)\\
&+\sum_{S_k< \pi}P_{0k}u'(W_{1k})(S_k-\pi)\left(1-\delta\frac{u(W_{1k})-\EE_0[u(W_1)]}{\SS_0[u(W_1)]}\right)\\
&<(1+|2\alpha-1|)\sum_{S_k\geq\pi}P_{0k}u'(W_{1k})(S_k-\pi)+(1-|2\alpha-1|)\sum_{S_k<\pi}P_{0k}u'(W_{1k})(S_k-\pi)\\
&=  c_\alpha\EE_0[u'(W_1)(S-\pi)|S\geq \pi]+c'_\alpha \EE_0[u'(W_1)(S-\pi)|S< \pi],
\end{align*}
where $c_\alpha:=(1+|2\alpha-1|)\PP_0[S\geq \pi]$ and $c'_\alpha:=(1-|2\alpha-1|)\PP_0[S< \pi]$, and the inequality follows from Lemma \ref{L:condition}. Note that when $S<\pi$,  it holds that, for all $\theta>0$, $u'(W_1)(S-\pi)$ is increasing in $S$ and hence there exists a state $\hat{k}_-$ such that $S_{\hat{k}_-}<\pi$ and $\EE_0[u'(W_1)(S-\pi)|S< \pi]\leq u'(W_{1\hat{k}_{-}})(S_{\hat{k}_{-}}-\pi)<0$.  

We first show the desired limits when $\Theta=\mathbb{R}$. We get that, for all $\alpha\in(0,1)$, $c'_\alpha$ is positive and in fact
\begin{align*}
\underset{\theta\uparrow\infty}{\lim}V'_\alpha(\theta)&\leq c_\alpha\underset{\theta\uparrow\infty}{\lim}\EE_0[u'(W_1)(S-\pi)|S\geq \pi]+c'_\alpha(S_{\hat{s}_{-}}-\pi)  \underset{\theta\uparrow\infty}{\lim}u'(W_{1\hat{s}_{-}})\\
&\leq c_\alpha(\overline{S}-\pi)\underset{x\uparrow\infty}{\lim}u'(x)+c'_\alpha(S_{\hat{s}_{-}}-\pi)  \underset{x\uparrow-\infty}{\lim}u'(x)=-\infty.
\end{align*}
When, $\alpha=0$, we may consider a positive $\tilde{\alpha}$ sufficiently close to zero such that $\tilde{d}=d(1-2\tilde{\alpha})$ satisfies Assumption \ref{A:condition for c} and conclude the same result. On the other hand, when $\alpha=1$, we directly get that $V_1(\theta)<V_\alpha(\theta)$ for all $\theta\in\Theta$ and $\alpha\in(1/2,1)$. Also, since $V_\alpha$ is strictly concave for all $\alpha\in[1/2,1]$ (thanks to Proposition \ref{P:concavity}), it holds that 
\[\underset{\theta\uparrow\infty}{\lim}V'_\alpha(\theta)=\underset{\theta\uparrow\infty}{\lim}\frac{V_\alpha(\theta)}{\theta},\qquad\forall\alpha\in[1/2,1].\]
Therefore, for $\alpha\in[1/2,1)$
\[\underset{\theta\uparrow\infty}{\lim}V'_1(\theta)=\underset{\theta\uparrow\infty}{\lim}\frac{V_1(\theta)}{\theta}\leq\underset{\theta\uparrow\infty}{\lim}\frac{V_\alpha(\theta)}{\theta}= \underset{\theta\uparrow\infty}{\lim}V'_\alpha(\theta)=-\infty.\]
The proof for the limit $\underset{\theta\downarrow-\infty}{\lim}V'_\alpha(\theta)=\infty$ is similar. Indeed, since we have shown that $\underset{\theta\uparrow\infty}{\lim}V'_\alpha(\theta)=-\infty$, we may consider the asset $\hat{S}:=-S$ at price $\hat{\pi}:=-\pi$ (note that we have not assumed positive values for the risky asset), and hence $V'_\alpha(\theta)$ can be written as 
\begin{align*}
\EE_0\left[ u'(w_0-\theta(\hat{S}-\hat{\pi}))(\hat{S}-\hat{\pi})\left(1-\delta\frac{u(w_0-\theta(\hat{S}-\hat{\pi}))-\EE_0[u(w_0-\theta(\hat{S}-\hat{\pi}))]}{\SS_0[u(w_0-\theta(\hat{S}-\hat{\pi}))]}\right) \right]
\end{align*}
which is equal to $-\hat{V}'_\alpha(-\theta)$, where $\hat{V}_\alpha$ denotes the value function with risky asset $\hat{S}$ at price $\hat{\pi}$. Therefore, 
\[\underset{\theta\downarrow-\infty}{\lim}V'_\alpha(\theta)=-\underset{\theta\downarrow-\infty}{\lim}\hat{V}'_\alpha(-\theta)=-\underset{\theta\uparrow\infty}{\lim}\hat{V}'_\alpha(\theta)=+\infty.\]

We now consider the case where $\Theta=\Theta_{w_0}(\pi )$. The proof follows similar arguments. We recall the inequality 
\[V'_\alpha(\theta)<c_\alpha\EE_0[u'(W_1)(S-\pi)|S\geq \pi]+c'_\alpha \EE_0[u'(W_1)(S-\pi)|S< \pi],\]
which can be rewritten as 
\begin{align*}
V'_\alpha(\theta)<& c_\alpha\EE_0[u'(W_1)(S-\pi)|S\geq \pi]+(1-|2\alpha-1|)\sum_{\underline{S}<S<\pi}P_{0s}u'(W_{1s})(S_s-\pi)\\
+& (1-|2\alpha-1|)P_{\underline{S}}u'(w_0+\theta(\underline{S}-\pi))(\underline{S}-\pi).
\end{align*}
Consider any $\alpha\in(0,1)$ and take the limit on both sides as $\theta\uparrow\overline{\Theta}=w_0/(\underline{S}-\pi)$ to get 
\begin{align*}
\underset{\theta\uparrow\overline{\Theta}}{\lim}V'_\alpha(\theta) &\leq c_\alpha\EE_0\left[\left. u'\left(w_0\frac{S-\underline{S}}{S-\pi}\right)(S-\pi)\right\vert S\geq \pi\right]\\
 &+(1-|2\alpha-1|)\sum_{\underline{S}<S<\pi}P_{0s}u'\left(w_0\frac{S_s-\underline{S}}{S_s-\pi}\right)(S_s-\pi)\\
&+ (1-|2\alpha-1|)P_{0\underline{S}}(\underline{S}-\pi)\, \underset{x\downarrow 0}{\lim} u'(x)=-\infty,
\end{align*}
since the first two terms of the right hand side are finite numbers. As before, when $\alpha=0$ we consider  positive $\tilde{\alpha}$ sufficiently close to zero such that $\tilde{d}=d(1-2\tilde{\alpha})$ satisfies Assumption \ref{A:condition for c} and conclude the same result. The case when $\alpha=1$ needs a proof only when $\underset{x\downarrow 0}{\lim}\,u(x)=-\infty$, since otherwise $V'_1(\theta)=\EE_{\PP^*}\left[ u'(W_1)(S-\pi)\right]$ for some measure $\PP^*(\theta)$ whose limit as $\theta\uparrow\overline{\Theta}$ is well-defined. 
We will develop this proof for any $\alpha\in[1/2,1]$. Due to the concavity of $V_\alpha$ for all $\alpha\in[1/2,1]$ (see Proposition \ref{P:concavity}) we get that 
\[\underset{\theta\uparrow\overline{\Theta}}{\lim}V'_\alpha(\theta)\leq\underset{\theta\uparrow\overline{\Theta}}{\lim} \frac{V_\alpha(\theta)-V_\alpha(0)}{\theta}=\frac{\lim_{\theta\uparrow\overline{\Theta}}V_\alpha(\theta)-V_\alpha(0)}{\overline{\Theta}}.\]
Now, denoting $u_{\underline{S}}:=u(w_0+\theta(\underline{S}-\pi))$  \eqref{eq:V_a nice} yields 
\begin{align*}
V_{\alpha}(\theta) &= P_{0\underline{S}}u_{\underline{S}}+ (1-P_{0\underline{S}})\EE_0[u(W_1)|S>\underline{S}]-\delta\SS_0[u(W_1)].
\end{align*}
Since
\begin{align*}
\SS^2_0[u(W_1)]&=(1-P_{0\underline{S}})\left[P_{0\underline{S}}(u_{\underline{S}}-\EE_0[u(W_1)|S>\underline{S}])^2+\SS_0^2[u(W_1)|S>\underline{S}] \right],
\end{align*}
we readily get that 
\begin{align*}
 V_{\alpha}(\theta) &\leq P_{0\underline{S}}u_{\underline{S}}+ (1-P_{0\underline{S}})\EE_0[u(W_1)|S>\underline{S}]\\
&-\delta\sqrt{(1-P_{0\underline{S}})P_{0\underline{S}}}\left|u_{\underline{S}}-\EE_0[u(W_1)|S>\underline{S}]\right|-\delta\sqrt{1-P_{0\underline{S}}}\SS_0[u(W_1)|S>\underline{S}]. 
\end{align*}
For positive $\theta$, it holds that $\EE_0[u(W_1)|S>\underline{S}]>u_{\underline{S}}$, and hence
\begin{align*}
 V_{\alpha}(\theta) &\leq u_{\underline{S}}\left(P_{0\underline{S}}+\delta\sqrt{(1-P_{0\underline{S}})P_{0\underline{S}}}\right)+ (1-P_{0\underline{S}})\EE_0[u(W_1)|S>\underline{S}]\\
&-\delta\sqrt{(1-P_{0\underline{S}})P_{0\underline{S}}}\EE_0[u(W_1)|S>\underline{S}]-\delta\sqrt{1-P_{0\underline{S}}}\SS_0[u(W_1)|S>\underline{S}].
\end{align*}
Assumption \ref{A:condition for c} implies that $\underline{c}:=P_{0\underline{S}}+\delta\sqrt{(1-P_0{\underline{S}})P_{0\underline{S}}}>0$. Recalling that $\overline{\Theta}=w_0/(\pi-\underline{S})$, we calculate
\begin{align*}
&\underset{\theta\uparrow\bar{\Theta}}{\lim}V_{\alpha}(\theta)\leq\underline{c}\,\underset{\theta\uparrow\bar{\Theta}}{\lim}\,u(w_0+\theta(\underline{S}-\pi))\\
&+\underset{\theta\uparrow\bar{\Theta}}{\lim}\left( (1-P_{0\underline{S}})\EE_0[u(W_1)|S>\underline{S}]
-\delta\sqrt{1-P_{0\underline{S}}}\left(\sqrt{P_{0\underline{S}}}\EE_0[u(W_1)|S>\underline{S}]+\SS_0[u(W_1)|S>\underline{S}]\right)\right)\\
&=-\infty,
\end{align*}
since the second limit is finite. This completes the proof of the first limit on \eqref{eq:large theta}.

Finally, the proof of the second limit of \eqref{eq:large theta} when $\Theta=\Theta_{w_0}(\pi )$ follows similar arguments and hence is omitted. 
\end{proof}

\subsection{Proposition \ref{P:demand_function}}
\begin{proof}
We first develop the proof for the case where $\Theta=\mathbb{R}$. 
For the first item, fix a price $\pi \in \left(\underline{S},\overline{S}\right)$. From Proposition \ref{P:concavity}, we get that $V'_\alpha$ is continuous and strictly decreasing for all $\theta\neq 0$. Therefore, $V_\alpha'$ has at most one root. 

For the existence of a unique root it is enough to show that 
\begin{equation}\label{eq:limits}
\underset{\theta\uparrow+\infty}{\lim}\,V_\alpha'(\theta)<0<\underset{\theta\downarrow-\infty}{\lim}\,V_\alpha'(\theta).
\end{equation}
For this, due to the strict concavity of $V_\alpha$, we have that  
\[\underset{\theta\uparrow+\infty}{\lim}\,V_\alpha'(\theta)=\underset{\theta\uparrow+\infty}{\lim}\,\frac{V_\alpha(\theta)}{\theta}.\]
Then,
\begin{align*}
\underset{\theta\uparrow+\infty}{\lim}\,\frac{V_\alpha(\theta)}{\theta}\leq\underset{\theta\uparrow+\infty}{\lim}\,\frac{\EE_0[u(W^{ \theta})]}{\theta}\leq\underset{\theta\uparrow+\infty}{\lim}\,\frac{u(w_0+\theta(\bar{S}-\pi))}{\theta}=\underset{\theta\uparrow+\infty}{\lim}\,u'(\theta)=0.
\end{align*}
Similarly,
\[\underset{\theta\downarrow-\infty}{\lim}\,\frac{V_\alpha(\theta)}{\theta}=-\underset{\theta\uparrow+\infty}{\lim}\,\frac{V_\alpha(-\theta)}{\theta}=-\underset{\theta\uparrow+\infty}{\lim}\,(V_\alpha(-\theta))'=\underset{\theta\downarrow-\infty}{\lim}\,V'_\alpha(\theta).\]
Now, 
\begin{align*}
\underset{\theta\downarrow-\infty}{\lim}\,\frac{V_\alpha(\theta)}{\theta}&\geq\underset{\theta\downarrow-\infty}{\lim}\,\frac{\EE_0[u(W^{ \theta})]}{\theta}\geq\underset{\theta\downarrow-\infty}{\lim}\,\frac{u(w_0+\theta(\underline{S}-\pi))}{\theta}=-\underset{\theta\uparrow+\infty}{\lim}\,\frac{u(w_0+\theta(\pi-\underline{S}))}{\theta}\\
&=-\underset{\theta\uparrow+\infty}{\lim}\,u'(\theta)=0.
\end{align*}
Since $V'_{\alpha}$ is strictly decreasing in $\RR$, we also get  strict inequalities in \eqref{eq:limits}.

As for the second item, we first show that $\tstar(\pi)=0$, for all $\pi \in[\upr,\opr]$. Let us first consider a price $\pi\in[\EE_0[S],\opr]$. From  Jensen's inequality and representation \eqref{eq:V_a nice}, we get that $\tstar(\EE_0[S])=0$. For any price $\pi>\EE_0[S]$, it holds that $\tstar(\pi)\leq 0$: Indeed, if we assume otherwise, by strict concavity of $V'$, $0=V'_\alpha(\tstar(\pi))<\lim_{\theta\downarrow 0}V'_\alpha(\theta)$. The latter is negative for any price $\pi>\EE_0[S]$ (due to by \eqref{eq:V'_limit+}), which yields a contradiction. 

We similarly show that for $\pi\leq \opr$, $\tstar(\pi)\geq 0$: Indeed if we assume otherwise, 
by strict concavity of $V'$ and \eqref{eq:V'_limit-}, we have \[0=V'_\alpha(\tstar(\pi))>\lim_{\theta\uparrow 0}V'_\alpha(\theta)=u'(w_0)(\EE_0[S]-\pi+\delta\SS_0[S])\geq 0,\]
a contradiction. Hence, for all $\pi\in[\EE_0[S],\opr]$, $\tstar(\pi)=0$. We work similarly to get that for all $\pi\in[\upr,\EE_0[S]]$, $\tstar(\pi)=0$. The next step is to show that for any $\pi>\opr$, $\tstar(\pi)<0$: Indeed, since we have shown that $\tstar(\pi)\leq 0$ for any price higher than $\EE_0[S]$, it is enough to show that $\tstar(\pi)\neq 0$. Assuming otherwise, we have that $\forall\theta<0$, $V_\alpha(\theta)<V_\alpha(0)$, which implies that $(V_\alpha(\theta)-V_\alpha(0))/\theta>0$. Taking the limit as $\theta\uparrow 0$ and recalling \eqref{eq:V'_limit-} give that $\underset{\theta\uparrow 0}{\lim}V'_{\alpha}(\theta)=u'(w_0)\left(\EE_0[S]-\pi+\delta\SS_0[S]\right)>0$, which contradicts the fact that $\pi$ is assumed higher than $\opr=\EE_0[S]+\delta\SS_0[S]$. Therefore, $\tstar(\pi)<0$ for any price $\pi>\opr$. It is left to observe that thanks to representation of FOC \eqref{eq:FOC} $\tstar(\pi)$ is continuous at $\opr$ and $\upr$.

As for the third item, for any price $\pi\leq \underline{S}$, optimal demand is positive (since $\pi<\upr\leq \underline{S}$) and from Proposition \ref{P:increasing} that $V_\alpha(\theta)$ is an increasing function of $\theta$. Similarly, for any price $\pi\geq \overline{S}$, optimal demand is negative (since $\underline{S}>\opr\geq  \pi$), and from Proposition \ref{P:increasing} that $V_\alpha(\theta)$ is a decreasing function of $\theta$.

Let us now consider the case $\Theta=\Theta_{w_0}(\pi)$ (i.e., utility functions that admit only positive wealth). We readily get that $\Theta=(\underline{\Theta},\overline{\Theta})$, where $\underline{\Theta}:=-w_0/(\overline{S}-\pi)$ and $\overline{\Theta}:=w_0/(\pi-\underline{S})$. We observe that the same proof holds for the second (portfolio inertia) and the third items. As for the first item of the Proposition, we consider initially the case of prices in $(\underline{S},\upr=\EE_0[S]-\delta\SS_0[S])$. For this, we get from \eqref{eq:V'_limit-} that  
\[\underset{\theta\uparrow 0}{\lim}V'_{\alpha}(\theta)>2u'(w_0)\delta\SS_0[S]>0,\]
which implies that for all $\theta<0$, it holds that $V'_\alpha(\theta)>V'_\alpha(0)>0$. For the existence of an optimal $\theta^*(\pi)\in\Theta$ for prices $p<\upr$ it is enough to show that there exists a positive $\theta^*$ for each $V'_\alpha(\theta)<0$. If we assume otherwise, then $V_\alpha$ is decreasing for all positive $\theta$ and hence $V_\alpha(\theta)\leq \underset{\theta\uparrow\overline{\Theta}}{\lim}\,V_\alpha(\theta)$. But then, 
\begin{align*}
\underset{\theta\uparrow\overline{\Theta}}{\lim}\,V_\alpha(\theta) &= \underset{\theta\uparrow\overline{\Theta}}{\lim}\{\EE_0[u(W^{ \theta})]-\delta\SS_0[u(W^{ \theta})]\}\\
&\leq \underset{\theta\uparrow\overline{\Theta}}{\lim}\,\EE_0[u(W^{ \theta})]=\underset{\theta\uparrow\overline{\Theta}}{\lim}\,\sum_{s\in D}p_s u(w_0+\theta(S_s-\pi)).
\end{align*}
To reach a contradiction, choose  $\theta$ sufficiently close to $\overline{\Theta}$, that is $\theta>\overline{\Theta}-\epsilon$ for some arbitrarily small $\epsilon>0$. We observe that for such a choice
\[w_0+\theta(\underline{S}-\pi)<\epsilon':=\epsilon(\pi-\underline{S}).\]
This means that there exist $\hat{\theta}\in\Theta\cap(0,\infty)$ sufficiently close to $\overline{\Theta}$, and at least one $\hat{s}\in D$ such that $w_0+\hat{\theta}(S_{\hat{s}}-\pi)\leq 0$. The latter implies that
\[\EE_0[u(W^{\hat{\theta}})]=p_{\hat{s}}u(w_0+\hat{\theta}(S_{\hat{s}}-\pi))+\sum_{s\neq \hat{s}}p_s u(w_0+\hat{\theta}(S_s-\pi))=-\infty,\]
a contradiction. 

The proof of the existence of a root of $V'_{\alpha}(\theta)$ for prices in $(\opr=\EE_0[S]+\delta\SS_0[S],\overline{S})$ follows similar steps. 
\end{proof}

\smallskip

\subsection{Proposition \ref{P:comparative_statics_delta}}

\begin{proof}
From Proposition \ref{P:demand_function}, we know that for prices $\pi$ higher than $\opr$ (resp.~lower than $\upr$), there exists a unique negative (resp.~positive) optimal $\theta^*$ that satisfies FOC \eqref{eq:FOC}. With a slight abuse of notation we may parametrize FOC with respect to $\alpha$ as
\begin{equation}\label{eq:FOC_for proof}
\EE_{\PP*(\alpha)}[u'(w_0+\theta^*(\alpha)(S-\pi))(S-\pi)]=0
\end{equation}
where 
\[\frac{\PP^*(\alpha)}{\PP_0}=1-\sqrt{d}(2\alpha-1) \frac{u(w_0+\theta^*(\alpha)(S-\pi))-\EE_0\left[u(w_0+\theta^*(\alpha)(S-\pi))\right]}{\SS_0[u(w_0+\theta^*(\alpha)(S-\pi))]}.\]
We now take the partial derivative of \eqref{eq:FOC_for proof} and solve with respect to $\partial\theta^*(\alpha)/\partial\alpha$ to get
\begin{equation}\label{eq:derivative of theta wrt alpha}\frac{\partial\theta^*(\alpha)}{\partial\alpha}=2\sqrt{d}\frac{\Cov(u(W_1^*),u'(W_1^*)(S-\pi))/\SS_0[u(W_1^*)]}{\EE_{\PP^*}[u''(W_1^*)(S-\pi)^2]-\delta(1-\rho^2)(\SS^2_0[u'(W_1^*)(S-\pi)]/\SS_0[u(W_1^*)])},
\end{equation}
where $\rho$ stands for the correlation coefficient between $u(W_1^*)$ and $u'(W_1^*)(S-\pi)$. Note that since $\delta\geq 0$, the denominator is negative and hence the sign of $\partial\theta^*(\alpha)/\partial\alpha$ is determined by the numerator's sign. In fact, the denominator is equal to $V''_\alpha(\theta^*)$, see \eqref{eq:horrorderivative}.

Let us now consider a price $\pi\in(\underline{S},\upr)$, which implies that $\theta^*$ is positive. We argue that if relative risk aversion is bounded from above by one, $\Cov(u(W_1^*),u'(W_1^*)(S-\pi))$ is positive. Indeed, for positive positions, $u(W_1)$ is increasing with respect to $S$. On the other hand $u'(W_1)(S-\pi)$ is also increasing as a function of $S$, since
\begin{align*}
\frac{\partial u'(W_1)(S-\pi)}{\partial S}&=u''(W_1)\theta(S-\pi)+u'(W_1)= u''(W_1)(W_1-w_0)+u'(W_1).
\end{align*} 
But, $u''(W_1)(W_1-w_0)+u'(W_1)>0$ is equivalent to 
\begin{align*}
\frac{u''(W_1)W_1}{u'(W_1)}- \frac{u''(W_1)w_0}{u'(W_1)}+1 &>0
\end{align*}
which is, in turn, equivalent to 
\[-R(W_1)- \frac{u''(W_1)w_0}{u'(W_1)} +1>0\quad \Leftrightarrow \quad R(W_1) <-\frac{u''(W_1)w_0}{u'(W_1)}+1,\]
where $R(x):=-u''(x)x/u'(x)$ stands for the relative risk aversion of $u$. The latter inequality holds true under the assumption that $R(x)\leq 1$, for all $x$. Hence, $\theta^*$ is a decreasing function of $\alpha$. 

On the other hand, for any price $\pi\in(\opr,\bar{S})$, we have that $\theta^*$ is negative. This implies that $u$ is a decreasing function of $S$, and since $u'(W_1^*)(S-\pi)$ is increasing (as long as $R(x)<1$), we have that $\Cov(u(W_1^*),u'(W_1^*)(S-\pi))$ is negative and hence $\theta^*$ becomes an increasing function of $\alpha$.  
%
%

\end{proof}

\subsection{Proposition \ref{P:seeker}}
\begin{proof}
For the first item, we get from  Jensen's inequality  that for all $\theta>0$ and when $\pi\geq\opr$ it holds 
\begin{align*}
V_\alpha(\theta)&\leq \alpha\underset{P\in\XX}{\min}\{u(w_0+\theta(\EE_{P}[S]-\pi))\}+(1-\alpha)\underset{P\in\XX}{\max}\{u(w_0+\theta(\EE_{P}[S]-\pi))\}\\
&=\alpha u\left(w_0+\theta(\underset{P\in\XX}{\min}\{\EE_{P}[S]\}-\pi)\right)+(1-\alpha)u\left(w_0+\theta(\underset{P\in\XX}{\max}\{\EE_{P}[S]\}-\pi)\right)\\
&=\alpha u\left(w_0+\theta(\EE_{0}[S]-\sqrt{d}\SS_0[S]-\pi)\right)+(1-\alpha)u\left(w_0+\theta(\EE_{0}[S]+\sqrt{d}\SS_0[S]-\pi)\right)\\
&\leq u(w_0+\theta(\opr-\pi))<u(w_0)=V_\alpha(0).
\end{align*}
Similar calculations yield that, for all $\theta<0$ and when $\pi<\upr$,  
$$V_\alpha(\theta)\leq u(w_0+\theta(\upr-\pi))<u(w_0)=V_\alpha(0),$$
which proves the second item. 

For the third one, consider price $\pi\in[\EE_0[S],\overline{\pi})$. We get from \eqref{eq:V'_limit+} that 
\[\underset{\theta\downarrow 0}{\lim}V'_{\alpha}(\theta)=u'(w_{0})\left(\opr-\pi\right)>0.\]
This means that for positive $\theta$ close to zero, $V_{\alpha}$ is strictly increasing and therefore, a positive locally optimal demand exists if there exists at least one $\theta^*_+>0$, such that $V'_{\alpha}(\theta^*_+)=0$. Assuming otherwise means that $V'_{\alpha}(\theta)>0$, for all $\theta>0$. However, this will contradict the result of Lemma \ref{lem:large theta} and in particular the fact that $\lim_{\theta\uparrow\overline{\Theta}}V'_{\alpha}(\theta)=-\infty$. 

For prices in $[\EE_0[S],\overline{\pi})$ we have a negative demand too, however. Indeed, we get from \eqref{eq:V'_limit-}
that \[\underset{\theta\uparrow 0}{\lim}V'_{\alpha}(\theta)=u'(w_{0})\left(\upr-\pi\right)<u'(w_{0})\left(\EE_0[S]-\pi\right)\leq 0.\]
This means that for negative $\theta$ close to zero, $V_{\alpha}$ is strictly decreasing and therefore, a negative locally optimal demand exists if there exists at least one $\theta^*_->0$, such that $V'_{\alpha}(\theta^*_-)=0$. Assuming otherwise means that $V'_{\alpha}(\theta)<0$, for all $\theta<0$. However, this will contradict the result of Lemma \ref{lem:large theta} and in particular the fact that $\lim_{\theta\downarrow\underline{\Theta}}V'_{\alpha}(\theta)=+\infty$. 

For the last item, we consider a price $\pi\in(\underline{\pi},\EE_0[S]]$ and proceed similarly. We have from \eqref{eq:V'_limit-} that $\underset{\theta\uparrow 0}{\lim}V'_{\alpha}(\theta)=u'(w_{0})\left(\upr-\pi\right)<0.$
This means that for negative $\theta$ close to zero, $V_{\alpha}$ is strictly decreasing and therefore, the second item holds, if there exists at least one $\theta^*_-<0$, such that $V'_{\alpha}(\theta^*_-)=0$. Assuming otherwise means that $V'_{\alpha}(\theta)<0$, for all $\theta<0$. However, this will contradict the result of Lemma \ref{lem:large theta} and in particular the fact that $\lim_{\theta\downarrow\underline{\Theta}}V'_{\alpha}(\theta)=+\infty$.

Also for prices in $(\underline{\pi},\EE_0[S]]$ we have a positive demand. Indeed, we get from \eqref{eq:V'_limit+}
that \[\underset{\theta\downarrow 0}{\lim}V'_{\alpha}(\theta)=u'(w_{0})\left(\opr-\pi\right)>u'(w_{0})\left(\EE_0[S]-\pi\right)\geq 0.\]
This means that for positive $\theta$ close to zero, $V_{\alpha}$ is strictly increasing and therefore, a positive locally optimal demand exists if there exists at least one $\theta^*_+>0$, such that $V'_{\alpha}(\theta^*_+)=0$. Assuming otherwise means that $V'_{\alpha}(\theta)>0$, for all $\theta>0$. However, this will contradict the result of Lemma \ref{lem:large theta} and in particular the fact that $\lim_{\theta\uparrow\overline{\Theta}}V'_{\alpha}(\theta)=-\infty$. 

\end{proof}

\subsection{Proposition \ref{P:comparative_statics_delta_seeker}}

\begin{proof}
The proof follows the steps of the proof of Proposition \ref{P:comparative_statics_delta}. Indeed, for any local optimal demand the both FOC \eqref{eq:FOC} and representation \eqref{eq:derivative of theta wrt alpha} hold even when $\alpha\in[0,1/2)$. Again, if the relative risk aversion is bounded from below by one, the numerator of \eqref{eq:derivative of theta wrt alpha} is positive (for positive demands), while the denominator is equal to $V''_\alpha(\theta^*)$, which is negative for the local maximum demand. 
\end{proof}
\subsection{Proposition \ref{p:binomial_seeker}}
\begin{proof}
We state the proof of the first item, with the second being analogous.
Based on representation \eqref{eq:V_bin_new}, we first note that 
$\lim_{\theta\downarrow 0}V'_{\alpha}(\theta)=(V^+_{\alpha})'(0)$. Hence, the FOC for problem $\max_{\theta\geq 0}V_\alpha(\theta)$ is
\[(V^+_{\alpha})'(\theta)\leq 0,\quad\theta\geq 0\quad\text{and}\quad\theta\cdot (V^+_{\alpha})'(\theta)=0.\]
For price $\pi<\opr$, we have from \eqref{eq:V'_limit+} that $(V^+_{\alpha})'(0)=u'(w_0)(\opr-\pi)>0$. Hence, the  FOC  may hold only if $\hat{\theta}\cdot(V^+_{\alpha})'(\hat{\theta})=0$ for some positive $\hat{\theta}$. Since, probability measure $P_+$ does not depend on $\theta$, $(V^+_{\alpha})'(\hat{\theta})=0$ is equivalent to 
\[\frac{\EE_+[u'(W^{\hat{\theta}})S]}{\EE_+[u'(W^{\hat{\theta}})]}=\pi.\]
Since we work in a forward market,  the binomial model is complete,  and the second fundamental theorem of asset pricing guarantees that for every price $\pi\in(\underline{S},\overline{S})$, there is a unique probability measure $\QQ(\pi)$ such that $\EE_{\QQ(\pi)}[S]=\pi$. Therefore, the optimal $\hat{\theta}$ is the one that satisfies $u'(W^{\hat{\theta}})=\epsilon \QQ(\pi)/P_+,$
for some  constant $\epsilon >0$, or equivalently
\begin{equation}\label{eq:FOC for binomial}
W^{\hat{\theta}}=I\left(\epsilon \frac{\QQ(\pi)}{P_+}\right),
\end{equation}
where $I:=(u')^{-1}$ is a well-defined function (due to the strict concavity of $u$). Market completeness implies the existence of $\hat{\theta}\in\reals$ that solves \eqref{eq:FOC for binomial}, for each $\epsilon$. The exact choice of $\epsilon $  then follows from the monotonicity of function $I$
and the equation $\EE_{\QQ(\pi)}[W^{\hat{\theta}}]=w_0=\EE_{\QQ(\pi)}[I(\epsilon \QQ(\pi)/P_+]$. Furthermore, $\hat{\theta}$ is indeed positive since $V^+_{\alpha}(\theta)$ is strictly concave (hence its derivative is decreasing) and $(V^+_{\alpha})'(0)>0$. Finally, strict concavity of $V^+_{\alpha}(\theta)$ yields that the solution of the aforementioned FOC is the unique maximizer of the problem $\max_{\theta\geq 0}V_\alpha(\theta)$.

On the other hand, if $\pi\geq\opr$, we have that $\lim_{\theta\downarrow 0}V'_{\alpha}(\theta)=(V^+_{\alpha})'(0)\leq 0$ which implies that the optimal $\hat{\theta}=0$, since $(V^+_{\alpha})'(\theta)$ is strictly increasing.
\end{proof}

\subsection{Theorem \ref{Thm:equilibrium}}
\begin{proof}
The existence of an equilibrium follows from Proposition \ref{P:demand_function}, in particular the uniqueness of demand $\tstar_i(\pi)$, its continuity and its range. In particular, if we assume without loss of generality that \eqref{eq:condition for equilibrium} holds for investor 1 and 2, such that $ \opr_1<\upr_2$, then for prices $\pi>\opr_1$, $\theta_1^*(\pi)$ is negative (the first investor wants for sell), while for prices $\pi<\upr_2$, $\theta_2^*(\pi)$ is positive (the second investor wants for buy). Then, the existence of a market-clearing equilibrium follows by item iii. of Proposition \ref{P:demand_function}. 
\end{proof}

\subsection{Theorem \ref{P:Second-best_binomial}}
\begin{proof}
Let $\opr_2=\underset{i\in\II,\,i> 1}{\max}\,\opr_i$ and assume that $\opr_1>\opr_2$. According to Proposition \ref{p:binomial_seeker}, for all prices $\underline{S}<\pi<\opr$, $\tilde{\theta}_1(\pi)=\underset{\theta\geq 0}{\argmax}V_{\alpha_1}(\theta)$ is strictly positive. On the other hand, Proposition \ref{P:demand_function} states that demand function for the second investor is strictly negative for prices higher than $\pi>\opr_2$. Again from Proposition \ref{P:demand_function} (item iii.), we get that the range of $\theta_2^*(\pi)$ is $(\underline{\Theta},0)$ for prices $\pi>\opr_2$. This implies that there exists a price $\tilde{\pi}\in(\opr_2,\opr_1)$, such that $\tilde{\theta}_1(\tilde{\pi})=-\theta^*(\tilde{\pi})$. 
Similarly, we show that there exists a second-best Pareto-optimal price when $\upr_1<\underset{i\in\II,\,i> 1}{\min}\,\upr_i$.  
\end{proof}

\subsection{Proof of Proposition \ref{P:Local_Second-best}}
\begin{proof}
Without loss of generality, assume that $\opr_2=\underset{i\in\II,\,i> 1}{\max}\,\opr_i$ and $\opr_1>\opr_2$. We first fix a price $\pi\in(\opr_2,\opr_1)$. According to Proposition \ref{P:demand_function}, there exists a unique optimal demand for the second investor $\theta_2^*(\pi)<0$. On the other hand, we get from \eqref{eq:V'_limit+} that 
\[\underset{\theta\downarrow 0}{\lim}V'_{\alpha_1}(\theta)=u_1'(w_{1})\left(\opr_1-\pi\right)>0,\]
where $w_1$ is the initial wealth of the ambiguity seeker. Hence, for positive $\theta$ close to zero, $V_{\alpha_1}$ is increasing. Now, we have two cases. The first is that $V_{\alpha_1}$ is increasing up to $-\theta_2^*(\pi)>0$. At that case, price $\pi$ is a non-zero local second-best Pareto optimal equilibrium price for the subset $\Theta_r(\pi):=[\theta_2^*(\pi),-\theta_2^*(\pi)]$ since
\[-\theta_2^*(\pi)=\underset{\theta\in\Theta_r(\pi)\cap[0,\overline{\Theta}_1)}{\argmax}\, V_{\alpha _1}(\theta).\]
The second case is that $V_{\alpha_1}$ is not increasing up to $-\theta_2^*(\pi)>0$. This means that there exists a position $\theta_1^*(\pi)\in(0,-\theta_2^*(\pi))$, such that $V'_{\alpha _1}(\theta_1^*(\pi))=0$ and   
 \[\theta_1^*(\pi)=\underset{\theta\in(0,-\theta_2^*(\pi))}{\argmax}\, V_{\alpha _1}(\theta).\]
On the other hand, owing to its strict concavity, $V_{\alpha_2}$ is decreasing for all $\theta>\theta^*_1(\pi)$ and hence 
\[-\theta_1^*(\pi)=\underset{\theta\in[-\theta_1^*(\pi),\theta_1^*(\pi)]}{\argmax}\, V_{\alpha _2}(\theta).\]
Therefore, in this case price $\pi$ is a non-zero local second-best Pareto optimal equilibrium price  for the subset $[-\theta_1^*(\pi),\theta_1^*(\pi)]$. 
\end{proof}

\section{Differentiability at zero}\label{sec:diffatzero}
We consider the case of one asset (i.e., $K=1$) and $\alpha\geq 1/2$. Since, $u$ is twice differentiable, we get that $V_\alpha$ is continuous for all $\theta\in\Theta\subseteq\mathbb{R}$. Now, for any $\theta\in\Theta\setminus\{0\}$ we calculate
\begin{align}\label{eq:derivative}
\nonumber V_\alpha'(\theta)&=\EE_{0}[u'(W^{ \theta})(S-\pi)]-\sqrt{d}(2\alpha-1)(\SS_0[u(W^{ \theta})])' \\
				 &=\EE_{0}[u'(W^{ \theta})(S-\pi)]-\sqrt{d}(2\alpha-1)\frac{\EE_0\left[(u(W^{ \theta})-\EE_0[u(W^{ \theta})])u'(W^{ \theta})(S-p)\right]}{\SS_0[u(W^{ \theta})]}.   
\end{align}
Consider first the case where $\theta>0$. From \eqref{eq:derivative}, we calculate
\begin{align*}
\underset{\theta\downarrow 0}{\lim}V'_{\alpha}(\theta)&=\underset{\theta\downarrow 0}{\lim}\EE_0[u'(W^{ \theta})(S-\pi)]-\delta \underset{\theta\downarrow 0}{\lim}\left(\SS_0[u(W^{ \theta})]\right)'\\
&=u'(w_0)\EE_0[S-\pi]-\delta\underset{\theta\downarrow 0}{\lim}\frac{\Cov_0\left(u(W^{ \theta}),u'(W^{ \theta})(S-\pi)\right)}{\SS_0[u(W^{ \theta})]},
\end{align*}
where $\delta:=\sqrt{d}(2\alpha-1)>0$. Note that for values of $\theta$ close to zero, it holds that 
\begin{equation}\label{eq:covariance}
\theta>0\qquad \Leftrightarrow \qquad \Cov_0\left(u(W^{ \theta}),u'(W^{ \theta})(S-\pi)\right)>0,
\end{equation}
since the derivatives of functions $x:\mapsto u(w_0+\theta x)$ and $x:\mapsto u'(w_0+\theta x)x$ have the same (resp.~different) sign when $\theta>0$ ($\theta<0$). This is very important, because it changes the sign of the second term of $V_\alpha'$ for $\theta$ close to zero. Using De l' Hospital rule, we get that 
\begin{align*}
L_+ &:=\underset{\theta\downarrow 0}{\lim}\left(\SS_0[u(W^{ \theta})]\right)'=\frac{\underset{\theta\downarrow 0}{\lim}\left(\Cov_0\left(u(W^{ \theta}),u'(W^{ \theta})(S-\pi)\right)\right)'}{L_+}=\frac{\left(u'(w_0)\right)^2\VV_0[S]}{L_+}.
\end{align*}
By taking into account \eqref{eq:covariance}, we get that $L_+=u'(w_0)\SS_0[S]$, and hence
\begin{equation}\label{eq:V'_limit+}
\underset{\theta\downarrow 0}{\lim}V'_{\alpha}(\theta)=u'(w_0)\left(\EE_0[S]-\pi-\delta\SS_0[S]\right).
\end{equation}
We work similarly with the limit from below.
\begin{align*}
\underset{\theta\uparrow 0}{\lim}V'_{\alpha}(\theta)&=\underset{\theta\uparrow 0}{\lim}\EE_0[u(W^{ \theta})(S-\pi)]-\delta \underset{\theta\uparrow 0}{\lim}\left(\SS_0[u(W^{ \theta})]\right)'\\
&=u'(w_0)\EE_0[S-\pi]-\delta\underset{\theta\uparrow 0}{\lim}\frac{\Cov_0\left(u(W^{ \theta}),u'(W^{ \theta})(S-\pi)\right)}{\SS_0[u(W^{ \theta})]},
\end{align*}
Using De l' Hospital rule, we get that 
\begin{align*}
L_- &:=\underset{\theta\uparrow 0}{\lim}\left(\SS_0[u(W^{ \theta})]\right)'=\frac{\underset{\theta\uparrow 0}{\lim}\left(\Cov_0\left(u(W^{ \theta}),u'(W^{ \theta})(S-\pi)\right)\right)'}{L_-}=\frac{\left(u'(w_0)\right)^2\VV_0[S]}{L_-}.
\end{align*}
By taking into account \eqref{eq:covariance}, we get that $L_-=-u'(w_0)\SS_0[S]$ and hence
\begin{equation}\label{eq:V'_limit-}
\underset{\theta\uparrow 0}{\lim}V'_{\alpha}(\theta)=u'(w_0)\left(\EE_0[S]-\pi+\delta\SS_0[S]\right).
\end{equation}
Therefore, $V_\alpha'$ is continuous at zero if and only if $c=1$ or $\alpha=1/2$.

\end{appendix}

\bibliographystyle{ecta}
\bibliography{master}

\end{document}